\documentclass[10pt]{article} % For LaTeX2e
\usepackage[accepted]{tmlr}
\usepackage{amsmath,amsfonts,bm}

\def\eqref#1{equation~\ref{#1}}
\def\1{\bm{1}}

\DeclareMathAlphabet{\mathsfit}{\encodingdefault}{\sfdefault}{m}{sl}
\SetMathAlphabet{\mathsfit}{bold}{\encodingdefault}{\sfdefault}{bx}{n}

\usepackage{hyperref}
\usepackage{url}

\usepackage{amsmath}
\usepackage{siunitx}
\usepackage{adjustbox}
\usepackage{makecell}
\usepackage{multirow}
\usepackage{float}
\usepackage{supertabular,booktabs}
\usepackage{longtable}
\usepackage{tabularx}
\usepackage{stfloats}
\usepackage{enumitem}
\usepackage{color}
\newtheorem{definition}{Definition}
\usepackage[ruled,vlined,linesnumbered]{algorithm2e}
\usepackage{xurl}
\usepackage{colortbl}
\usepackage{wrapfig}
\usepackage{caption}
\usepackage{subcaption}
\usepackage{amssymb}
\newcommand{\specialcell}[2][c]{%
    \begin{tabular}[#1]{@{}c@{}}#2\end{tabular}
}

\title{Unsupervised Network Embedding Beyond Homophily}

\author{\name Zhiqiang Zhong \email zhiqiang.zhong@uni.lu \\
      \addr University of Luxembourg
      \AND
      \name Guadalupe Gonzalez \email ggg17@ic.ac.uk \\
      \addr Imperial College London
      \AND
      \name Daniele Grattarola \email daniele.grattarola@epfl.ch \\
      \addr École Polytechnique Fédérale de Lausanne
      \AND
      \name Jun Pang \email jun.pang@uni.lu\\
      \addr University of Luxembourg
}
\def\month{12}  % Insert correct month for camera-ready version
\def\year{2022} % Insert correct year for camera-ready version
\def\openreview{\url{https://openreview.net/pdf?id=sRgvmXjrmg}} % Insert correct link to OpenReview for camera-ready version

\begin{document}

\maketitle

%------------------------------------------------------------------------------
\begin{abstract} 
% Network embedding (NE) has emerged as a predominant technique for representing complex networks and benefited numerous tasks.
% However, most NE approaches implicitly rely on a homophily assumption or learn embeddings with the guidance of supervisory signals. 
% This paper focuses on one unexplored question: \textit{whether existing NE methods adapt well to networks with heterophily}, i.e., networks in which nodes with different labels tend to be linked, \textit{under no supervision}.
% An empirical study on the influence of the homophily ratio on the performance of existing unsupervised NE (UnSupNE) methods reveals that they are failing to adapt well to heterophily.
% Motivated by this observation, we formulate the UnSupNE task as an $r$-ego network discrimination problem and further develop the SELENE framework for learning over networks with homophily and heterophily.
% Specifically, we design a dual-channel feature embedding pipeline to discriminate $r$-ego networks from aspects of node attributes and structural information. 
% The introduced identity and network structural features further help to distinguish different sampled $r$-ego networks.
% Lastly, we employ negative-sample-free self-supervised learning objective functions to optimise the framework to learn intrinsic node embeddings.
% Our comprehensive empirical evaluations on both synthetic and real-world data with varying homophily ratios validate the effectiveness of SELENE with homophily and heterophily by up to $13.35\%$ clustering accuracy gain.

% NEW ABSTRACT\\
Network embedding (NE) approaches have emerged as a predominant technique to represent complex networks and have benefited numerous tasks. 
However, most NE approaches rely on a homophily assumption to learn embeddings with the guidance of supervisory signals, leaving the \textit{unsupervised} \textit{heterophilous} scenario relatively unexplored.
This problem becomes especially relevant in fields where a scarcity of labels exists. 
Here, we formulate the unsupervised NE task as an $r$-ego network discrimination problem and develop the SELENE framework for learning on networks with homophily and heterophily. 
Specifically, we design a dual-channel feature embedding pipeline to discriminate $r$-ego networks using node attributes and structural information separately. 
We employ heterophily adapted self-supervised learning objective functions to optimise the framework to learn intrinsic node embeddings. 
We show that SELENE's components improve the quality of node embeddings, facilitating the discrimination of connected heterophilous nodes.
Comprehensive empirical evaluations on both synthetic and real-world datasets with varying homophily ratios validate the effectiveness of SELENE in homophilous and heterophilous settings showing an up to $12.52\%$ clustering accuracy gain.

\end{abstract}
%------------------------------------------------------------------------------

%------------------------------------------------------------------------------
\section{Introduction} 
\label{sec:introduction}
Network embedding (NE) has become a predominant approach to finding effective data representations of complex systems that take the form of networks~\citep{CWPZ19}. 
NE approaches leveraging graph neural networks (GNNs)~\citep{DBV16,KW17,XHLJ19,ZLP222} have proven effective in (semi)-supervised settings and achieved remarkable success in various application areas, such as social, e-commerce, biology, and traffic networks~\citep{ZCZ20}. 
However, recent works demonstrated that classic supervised NE methods powered by GNNs, which typically follow a homophily assumption, have limited representation power on heterophilous networks~\citep{BWSS21,LHLHGBL21,CPLM21,ZIP22,ZLPZJY22}. 

\begin{figure*}[t]
\centering
\includegraphics[width=.8\linewidth]{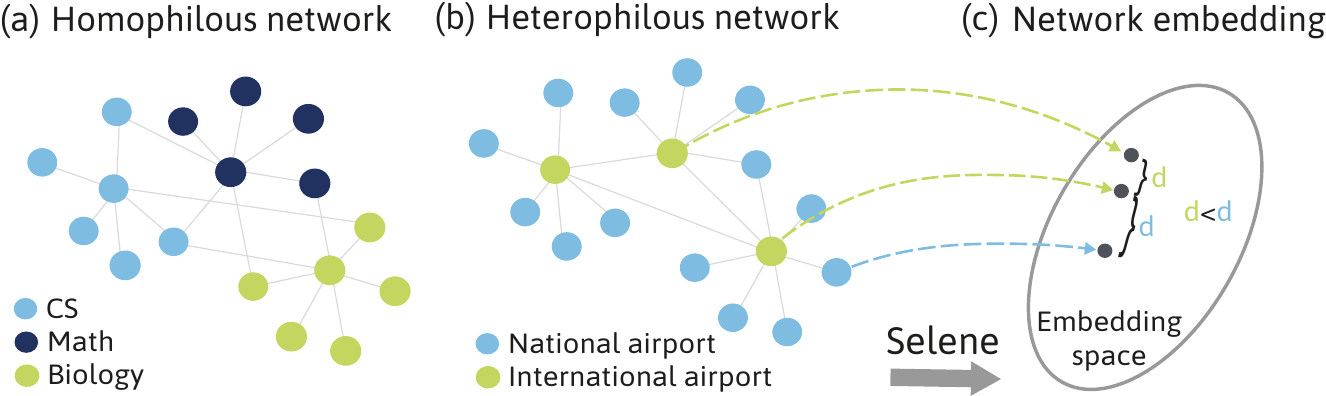}
\caption{
Example homophilous and heterophilous networks ((a): a citation network; (b): an airline transport network). 
(c) SELENE performs unsupervised network embedding capturing node attributes and structural information to address heterophily.
}
\label{fig:network_types_and_schema}
\vspace{-6mm}
\end{figure*}

% \textbf{Why is it interesting and important?}\\
Whereas similar nodes are connected in homophilous networks, the opposite holds for heterophilous networks in which connected nodes are likely from different classes (Figure~\ref{fig:network_types_and_schema}-(a-b)). 
For instance, people tend to connect with people of the opposite gender in dating networks~\citep{ZYZHAK20}
% different amino acid types are more likely to connect in protein structures~\citep{ZYZHAK20}; 
and fraudsters are more likely to connect with customers than other fraudsters in online transaction networks~\citep{PCWF07}. 
% Additionally, in application areas such as biomedical problems, where scarcity of labels exists, the unsupervised setting is of high interest to generate representations for various downstream tasks or pre-training~\citep{LHZ21}.

% \textbf{Why is it hard? (E.g., why do naive approaches fail?)}\\

Traditional GNNs typically fail in heterophilous scenarios because they obtain representations by aggregating information from neighbours, acting as a low-frequency filter which generates indistinguishable node representations on heterophilous networks (Figure \ref{fig:example_hete_network})~\citep{BWSS21}.
% \textbf{Why hasn’t it been solved before? (Or, what’s wrong with previous proposed solutions? How does mine differ?)}
Recently, several GNN operators have been introduced to overcome the smoothing effect of traditional GNNs on heterophilous networks~\citep{ZLPZJY22}, however, they rely heavily on the (semi)-supervised setting.
% which encourages embeddings of nodes with similar properties to be similar, partially helping the model overcome the issue of heterophily.
\citet{WYJBL22} conduct theoretical analyses to find out the conditions that GNN models have no performance difference on synthetic homophilous/heterophilous datasets. 
\citet{MLST22} empirically characterise different heterophily conditions and identify the specific conditions that lead to worse GNN model performance under supervised settings. 
In contrast, the effectiveness of GNNs in unsupervised settings, i.e. learning effective heterophilous representations without any supervision, is relatively unexplored~\citep{XSYAWPL21}.
In application areas such as biomedical problems, where scarcity of labels exists, the unsupervised setting is of high interest to generate representations for various downstream tasks~\citep{LHZ22}.
\textbf{Current limitations.} 
We address the task of node clustering under an unsupervised heterophilous setting (Figure~\ref{fig:network_types_and_schema}-(b-c)) (for convenience, we will refer to unsupervised network embedding as NE in the remainder of the text).
In this scenario, the first unexplored question to investigate is \textbf{RQ1}: \textit{how do existing NE methods perform on heterophilous networks without supervision?}

We conduct an empirical study on $10$ synthetic networks with a variety of homophily ratios ($h$, Definition~\ref{def:homophily_ratio}) to investigate whether $h$ influences the node clustering performance of representative NE methods.
Our experimental results, summarised in Figure~\ref{fig:performance_competing_methods_diff_syn_data}, show that \textit{(i)} the performance of NE methods that utilise network structure, including heterophilous GNNs designed for supervised settings, decreases significantly when $h \to 0$; 
\textit{(ii)} the performance of a NE method that only relies on raw node attributes is not affected by changes of $h$, but it is outperformed by other NE methods when $h \to 1$. 
These two findings directly answer RQ1 and meanwhile raise another interesting and challenging question \textbf{RQ2:} \textit{could we design a NE framework that adapts well to both homophily and heterophily settings under no supervision?}

\textbf{Our approach.} 
Motivated by the limitations mentioned above, we approach the NE task as an $r$-ego network discrimination problem and propose the \underline{SEL}f-sup\underline{E}rvised \underline{N}etwork \underline{E}mbedding (SELENE) framework. 
Our empirical study suggests that both node attributes and local structure should be leveraged to obtain node embeddings.
% Consequently, we define the node attributes and local structure as two aspects that are helpful to discriminate node embeddings, 
Therefore, we summarise them into an $r$-ego network and use self-supervised learning (SSL) objective functions to optimise the framework to compute distinguishable node embeddings.
Such a design assumes nodes of the same class label share either similar node attributes or $r$-ego network structure.
Specifically, we propose a dual-channel embedding pipeline that encodes node attributes and network structure information in parallel (Figure~\ref{fig:model_architecture}).
Next, after revisiting representative NE mechanisms, we introduce identity, and network structure features to enhance the framework's ability to capture structural information and distinguish different sampled $r$-ego networks. 
Lastly, since network sampling strategies often implicitly follow homophily assumptions, i.e., ``positively sampling nearby nodes and negatively sampling the faraway nodes''~\citep{YDZYZT20}, we employ negative-sample-free SSL objective functions, namely, reconstruction loss and Barlow-Twins loss, to optimise the framework. 

\textbf{Extensive evaluation.}
We empirically evaluate our model and competitive NE methods on both synthetic and real networks covering the full spectrum of low-to-high homophily and various topics, including node clustering, node classification and link prediction. 
We observe that SELENE achieves significant performance gains in both homophily and heterophily in real networks, with an up to $12.52\%$ clustering accuracy gain. 
Our detailed ablation study confirms the effectiveness of each design component. 
In synthetic networks $h \in [0, 1)$, we observe that SELENE shows better generalisation.
\section{Related Work} 
\label{sec:related_work}
Node clustering, one of the most fundamental graph analysis tasks, is to group similar nodes into the same category \textit{without supervision}~\citep{NJW01,S07}. 
Over the past decades, many clustering algorithms have been developed and successfully applied to various real-world applications~\citep{BN01}. 
Recently, the breakthroughs in deep learning have led to a paradigm shift in the machine learning community, achieving great success on many important tasks, including node clustering. 
Therefore, deep node clustering has caught significant attention~\citep{PSYTTRSCI19,YFSH17}. 
The basic idea of deep node clustering is to integrate the objective of clustering into the powerful representation ability of deep learning. 
Hence learning an effective node representation is a prerequisite for deep node clustering. 
To date, network embedding (NE)-based node clustering methods have achieved state-of-the-art performance and become the \textit{de facto} clustering methods.

% \noindent
\textbf{Network embedding before GNNs.}
NE techniques aim at embedding the node attributes and structure of complex networks into low-dimensional node representations~\citep{CWPZ19}.
Initially, NE was posed as the optimisation of an embedding lookup table directly encoding each node as a vector. 
Within this group, several methods based on skip-grams~\citep{MSCCD13} have been proposed, such as DeepWalk~\citep{PAS14}, node2vec~\citep{GL16}, struc2vec~\citep{RSF17}, etc~\citep{TQWZYM15,QDMLWT18}.
Despite the relative success of these NE methods, they often ignore the richness of node attributes and only focus on the network structural information, which hugely limits their performance. 

% \smallskip\noindent
\textbf{Network embedding with GNNs.}
Recently, GNNs have shown promising results in modelling structural and relational data~\citep{WPCLZY21}. 
GNN models capture the structural similarity of nodes through a recursive message-passing scheme, using neural networks to implement the message, aggregation, and update functions~\citep{BHBSZMT18}.
The effectiveness of GNNs has been widely proven in (semi)-supervised settings, and they have achieved remarkable success in various areas~\citep{ZCZ20}.
Several approaches such as ChebNet~\citep{DBV16}, GCN~\citep{KW17}, GraphSAGE~\citep{HYL17}, CayleyNets~\citep{LMBB19}, GWNN~\citep{XSCQC19}, and GIN~\citep{XHLJ19} have led to remarkable breakthroughs in numerous fields in (semi-)supervised settings. 
However, their effectiveness for unsupervised NE is relatively unexplored. 
Recently, GNN-based methods for unsupervised NE such as DGI~\citep{VFHLBH19}, GMI~\citep{PHLZRXH20}, SDCN~\citep{BWSZLC20}, and GBT~\citep{BKC21} have been proposed, although they were primarily designed for homophilous networks.

% \smallskip\noindent
\textbf{Heterophilous network embedding.}
Recent works have also focused on NE for heterophilous networks and have shown that the representation power of GNNs designed for (semi-)supervised settings is greatly limited on heterophilous networks~\citep{ZLPZJY22}. 
Some efforts have been dedicated to generalising GNNs to heterophilous networks by introducing complex operations, such as coarsely aggregating higher-order interactions or combining the intermediate representations~\citep{ZYZHAK20,BWSS21,LHLHGBL21}.
\citet{WYJBL22} conduct theoretical analyses to determine the conditions that GNN models have no performance difference on synthetic homophilous/heterophilous datasets. 
\citet{MLST22} empirically characterise different heterophily conditions and identify the specific conditions that lead to worse GNN model performance under supervised settings. 
Nevertheless, these heterophilous GNNs heavily rely on supervisory information and hence cannot be applied to unsupervised settings (verified in Section~\ref{sec:experimental_investigation} and Section~\ref{subsec:experimental_results}).
Latterly, \citet{TLY22} introduce Neighborhood Wasserstein Reconstruction loss and a novel decoder to properly capture node attributes and graph structure to learn discriminate node representations. 

%------------------------------------------------------------------------------

%------------------------------------------------------------------------------
\section{Notation and Preliminaries}
\label{sec:preliminaries}
An unweighted network can be formally represented as $\mathcal{G}=(\mathcal{V}, \mathcal{E}, \mathbf{X})$, where $\mathcal{V}$ is the set of nodes and $|\mathcal{V}| = n$ is the number of nodes, $\mathcal{E} \subseteq \mathcal{V} \times \mathcal{V}$ is the set of edges, and $\mathbf{X} \in \mathbb{R}^{n \times \pi}$ represents the $\pi$-dimensional node attributes. 
We let $\mathcal{Y} = \{y_v\}$ be a set of class labels for all $v \in \mathcal{V}$. 
For simplicity, we summarise $\mathcal{E}$ with an adjacency matrix $\mathbf{A} \in \{0, 1\}^{n \times n}$.

\begin{wrapfigure}{r}{0.33\textwidth}
% \begin{figure}[!ht]
\centering
\includegraphics[width=1.\linewidth]{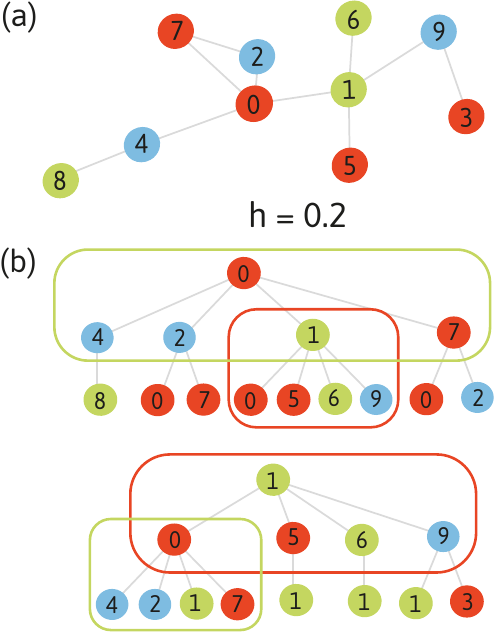}
\caption{(a) Network with $h = 0.2$; (b) the aggregation mechanism in traditional GNNs implicitly follows a homophily assumption as a result of duplicated aggregation trees for nearby nodes - as shown for trees rooted in nodes $v_0$ and $v_1$ (duplicates marked by same-coloured square).
% Different colours represent different node classes. 
% (a) An example network;
% (b) Two rooted $2$-layers aggregation trees of two connected nodes, i.e., $v_0$ and $v_1$. 
% Duplicated components of two rooted trees are marked with the same shadow colour. 
% (c) After ego network sampling and anonymity, two rooted aggregation trees have no duplicates. 
}
\label{fig:example_hete_network}
\vspace{-15mm}
% \end{figure}
\end{wrapfigure}

% \smallskip\noindent
\textbf{Problem setup}.
% This paper focuses on the unsupervised node clustering task. 
We setup the problem using the unsupervised node clustering task as an example. 
We first learn node representations $\textbf{Z}_{v} \in \mathbb{R}^{d}$ for all $v \in \mathcal{V}$, capturing both node attributes and local network structure.
Then, the goal is to infer the unknown class labels $y_{v}$ for all $v \in \mathcal{V}$ using a clustering algorithm on the learned node representation $\textbf{Z}_{v}$ (in this paper, we use the well-known \textit{K}-means algorithm~\citep{HW79}).
Note that, for convenience, in the following we refer to unsupervised NE simply as NE. 

% \smallskip\noindent
% \textbf{Definition 1.} \hspace{.1mm} (\textbf{$r$-hop Neighbourhood} $\mathcal{N}^{r}$).
\begin{definition}[$r$-hop Neighbourhood $\mathcal{N}^{r}$]
% We denote a general neighbourhood around ego node $v$, excluding $v$ (in case $\mathcal{G}$ has self-loops), as $\mathcal{N}_v$; and the corresponding neighbourhood including the ego node $v$ as $\widetilde{\mathcal{N}}_v$.
We denote the $r$-hop neighbourhood of node $v$ by $\mathcal{N}^{r}_v=\{v: d(u, v)\leq r\}$, where $d(u, v)$ is the shortest path distance between $u$ and $v$.
For the network shown in Figure~\ref{fig:example_hete_network}-(a), $\mathcal{N}^{1}_{v_0} = \{v_0, v_1, v_2, v_4, v_7 \}$. 
\end{definition}

% \smallskip\noindent
% \textbf{Definition 2.} \hspace{.1mm} (\textit{r}\textbf{-ego Network}.
\begin{definition}[\textit{r}-ego Network]
$\mathcal{G}_{r}(v)$)~\citep{ML12,QCDZYDWT20}
Let $\mathcal{N}^{r}_v \subseteq \mathcal{V}$ be the \textit{r}-ego neighbours of node $v$ in $\mathcal{G}$. 
Their corresponding \textit{r}-ego network is an induced sub-network of $\mathcal{G}$ defined as $\mathcal{G}_{r}(v)=\{\mathcal{N}^{r}_v, \mathcal{E}^{r}_{v}, \mathbf{X}^{r}_{v} \}$, where $\mathcal{E}^{r}_{v} := ((\mathcal{N}^{r}_v \times \mathcal{N}^{r}_v) \cap \mathcal{E})$. 
\end{definition}

% \smallskip\noindent
% \textbf{Definition 3.} \hspace{.1mm} (\textbf{Homophily Ratio} $h$).
\begin{definition}[Homophily Ratio $h$]
\label{def:homophily_ratio}
The homophily ratio $h$ of $\mathcal{G}$  describes the relation between node labels and network structure.
Recent works commonly use two measures of homophily, edge homophily ($h_{\it edge}$)~\citep{ZRRMLAK21} and node homophily ($h_{\it node}$)~\citep{PWCLY20}, which can be formulated as
$h_{\it edge} = \frac{|\{(u, v): (u, v) \in \mathcal{E} \wedge y_{u}=y_{v} \}|}{|\mathcal{E}|}, h_{\it node} = \frac{1}{|\mathcal{V}|}\sum_{v \in \mathcal{V}} \frac{|\{u: u \in \mathcal{N}^{1}_{v} \wedge y_{u}=y_{v}\}|}{|\mathcal{N}^{1}_{v}|}$
% \begin{equation}
% \label{eq:homophily}
%   \begin{aligned}
%   h_{\it edge} &= \frac{|\{(u, v): (u, v) \in \mathcal{E} \wedge y_{u}=y_{v} \}|}{|\mathcal{E}|} 
%   \qquad\qquad
%   h_{\it node} &= \frac{1}{|\mathcal{V}|}\sum_{v \in \mathcal{V}} \frac{|\{u: u \in \mathcal{N}^{1}_{v} \wedge y_{u}=y_{v}\}|}{|\mathcal{N}^{1}_{v}|}
%   \end{aligned}
% \end{equation}
Specifically, where $h_{\it edge}$ evaluates the fraction of edges between nodes with the same class labels;
$h_{\it node}$ evaluates the overall fraction of neighbouring nodes that have the same class labels.
In this paper, we focus only on edge homophily and denote it with 
% $h=h_{\it edge} = \frac{|\{(u, v): (u, v) \in \mathcal{E} \wedge y_{u}=y_{v} \}|}{|\mathcal{E}|}$.
$h=h_{\it edge}$.
Figure~\ref{fig:example_hete_network}-(a) shows an example network with $h=0.2$.
\end{definition}

%------------------------------------------------------------------------------

%------------------------------------------------------------------------------
\section{An Experimental Investigation}
\label{sec:experimental_investigation}
\begin{wrapfigure}{r}{0.4\textwidth}
% \begin{figure}[!ht]
\centering
% \subfloat[]{\includegraphics[width=1.\linewidth]{figures/plot_h_acc_syn.pdf}%
% \label{fig:h_acc_syn}}
% \hfil
% \subfloat[]{\includegraphics[width=1.\linewidth]{figures/plot_h_acc_products_syn.pdf}%
% \label{fig:h_acc_p-syn}}
% \caption{
% Node clustering accuracy of representative NE methods on synthetic datasets, i.e., Synthetic (a) and Synthetic-Products (b).
% % For detailed experimental results evaluated with multiple metrics we refer the reader to Section~\ref{subsec:experimental_results}
% }
\includegraphics[width=1.\linewidth]{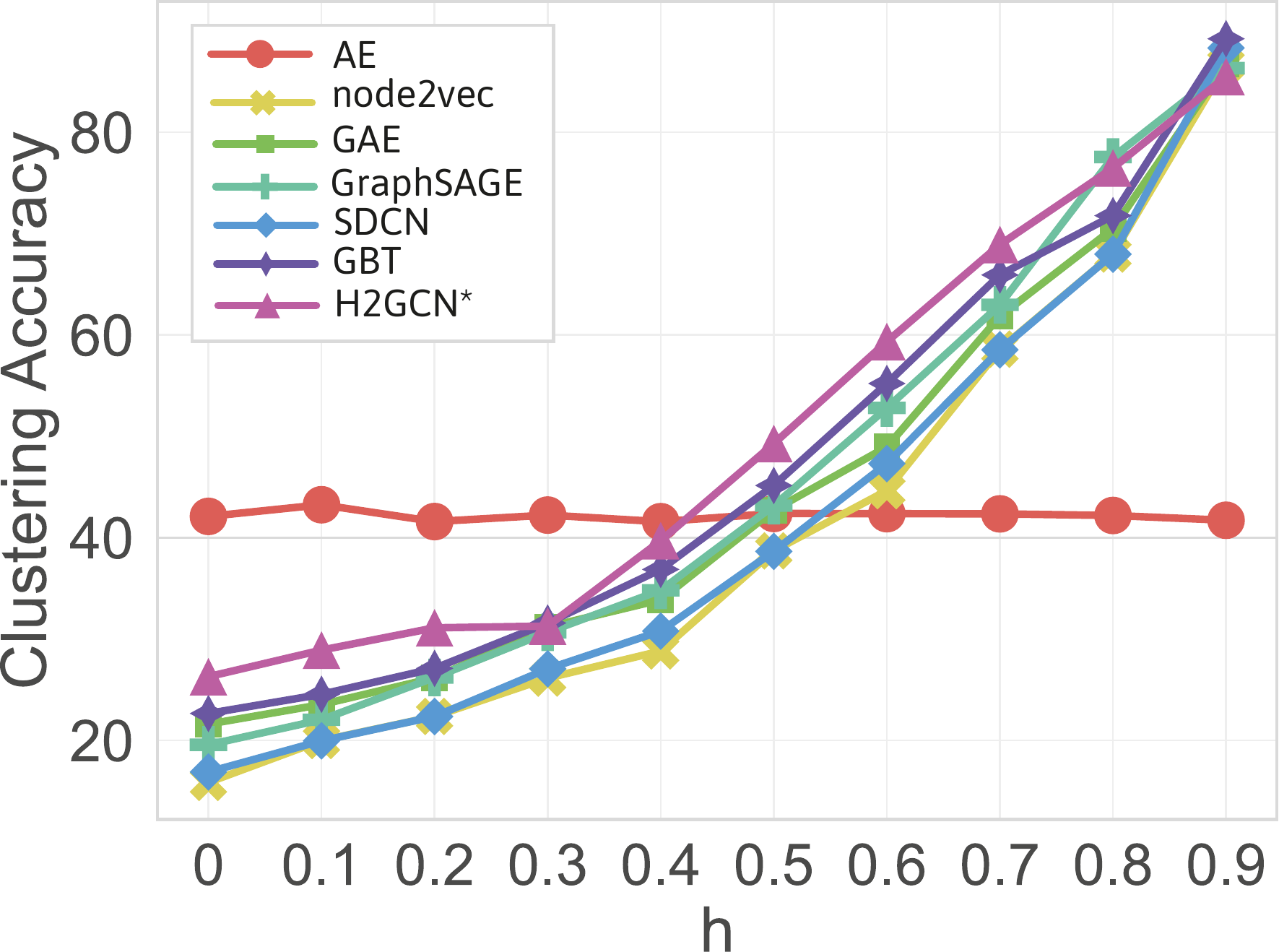}
\caption{
Node clustering accuracy of representative NE methods on synthetic networks.
}
\label{fig:performance_competing_methods_diff_syn_data}
\vspace{-5mm}
% \end{figure}
\end{wrapfigure}

In this section, we empirically analyse the performance of NE methods on $10$ synthetic networks with different homophily ratios ($h$).
The main goal is to investigate (\textbf{RQ1}): \textit{how do existing NE methods perform on heterophilous networks?} 
Specifically, we quantify their performance on the node clustering task on $10$ synthetic networks with $h \in [0, 0.1, \dots, 0.9]$. 
A detailed description of the synthetic datasets generation process can be found in Section~\ref{subsec:datasets}, and we refer the reader to Section~\ref{subsec:experimental_setup} for details on the experimental settings. 

Figure~\ref{fig:performance_competing_methods_diff_syn_data} illustrates that the clustering accuracy of representative NE methods that utilise network structure, i.e., node2vec~\citep{GL16}, GAE~\citep{KW16}, GraphSAGE~\citep{HYL17}, SDCN~\citep{BWSZLC20}, GBT~\citep{BKC21} and H2GCN$^{*}$~\citep{ZYZHAK20} show outstanding performance when $h \to 1$ but with the decrease of $h$,  their performance decreases significantly. 
The reason why existing NE methods using network structure fail only when $h \to 0$ is that most of them implicitly follow a homophily assumption, with specific objective function or aggregation mechanism designs.
For instance, \textit{(i)} objective functions of node2vec and GraphSAGE guide nodes at close distance to have similar representations and nodes far away to have different ones;
\textit{(ii)} the inherent aggregation mechanism of GAE, GraphSAGE, SDCN and GBT naturally assumes local smoothing~\citep{CLLLZS20} (which is mainly caused by the duplicated aggregation tree for nearby nodes as shown in Figure~\ref{fig:example_hete_network}-(b)), which translates into neighbouring nodes having similar representations. 

On the other hand, H2GCN$^{*}$, which is specifically designed for heterophilous networks, follows the same trend as the homophilous approaches due to the loss of supervisory signals in the network embedding task. 
The only method with stable performance across different values of $h$ is AE~\citep{HS06}, which is attributable to its reliance on raw node attributes only. AE exhibits an apparent advantage against all other models when $h < 0.5$.
This highlights the importance of considering node attributes in the design of NE approaches for networks with heterophily. 

% 
% Lastly, it is worth noting that struc2vec~\citep{RSF17}, which relies on network structure too, performs worse on the $10$ synthetic networks. 
% The reason is that struc2vec identifies the structure of ego networks by counting the degree of nodes of different hops. 
% The synthetic network generation process assigns node degree to each node with a normal distribution, i.e., nodes from different classes can have the same degree.
% Therefore, it is not able to identify the ego network's local structure.
% This suggests us that neural network-based models have better usability in capturing ego network structure patterns than manual defined statistical methods. 

%------------------------------------------------------------------------------

%------------------------------------------------------------------------------
\section{Network Embedding via \textit{r}-ego Network Discrimination}
\label{sec:approach}
\begin{figure*}[!t]
\centering
\includegraphics[width=.9\linewidth]{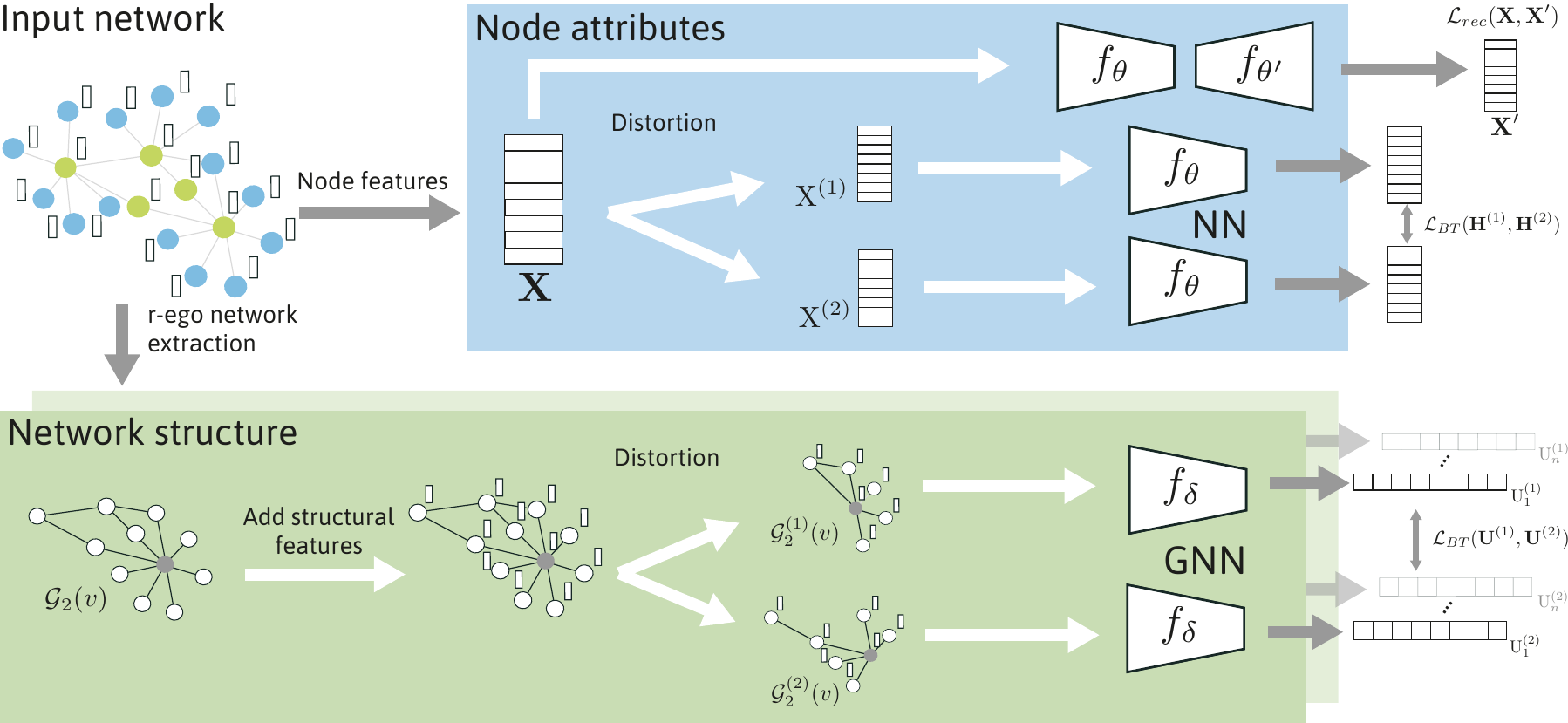}
\caption{
We propose a dual-channel feature embedding pipeline to learn node representations separately from node attributes and network structure. 
Node feature attribute embeddings are optimised using an autoencoder trained to reconstruct the input data $\mathcal{L}_{\it Rec}(\mathbf{X}, \widehat{\mathbf{X}})$, and refined through the contrastive loss $\mathcal{L}_{BT}(\textbf{H}^{(1)}, \textbf{H}^{(2)})$. 
The network structure encoder is optimised using $\mathcal{L}_{BT}(\mathbf{U}^{(1)}, \mathbf{U}^{(2)})$.
The final node representation is obtained by applying \textsc{COMBINE} to the generated node attribute representation $\mathbf{H}$ and network structure representation $\textbf{U}$.
% $\mathbf{X}$ and $\mathbf{A}$ are raw node attributes and adjacency matrix of the input network $\mathcal{G}$, respectively.
% $\mathbf{X}^{(1)}$ and $\mathbf{X}^{(2)}$ are two distorted node attribute matrix, $\widehat{\mathbf{X}}$ is the reconstructed node attribute matrix.
% Extracted $r$-ego network ($\mathcal{G}_{r}(v)$) of ego node $v$ will be anonymised to break its connection to neighbour nodes, and then distorted to two ego networks, i.e., $\mathcal{G}^{(1)}_{r}(v)$ and $\mathcal{G}^{(2)}_{r}(v)$. 
% Node attribute encoder is optimised by the reconstruction loss $\mathcal{L}_{\it Rec}(\mathbf{X}, \widehat{\mathbf{X}})$ and attribute Barlow-Twins loss $\mathcal{L}_{BT}(\mathbf{H}^{(1)}, \mathbf{H}^{(2)})$;
% network structure encoder is optimised by the network Barlow-Twins loss $\mathcal{L}_{BT}(\mathbf{U}^{(1)}, \mathbf{U}^{(2)})$.
% The final node representation obtained by applying \textsc{COMBINE} to the generated node attribute representation $\mathbf{H}$ and network structure representation $\textbf{U}$. 
}
\label{fig:model_architecture}
% \vspace{-3mm}
\end{figure*}

In this section, we formalise the main challenges of NE on heterophilous networks. 
To address these challenges, we present the \underline{SEL}f-sup\underline{E}rvised \underline{N}etwork \underline{E}mbedding (SELENE) framework.
Figure~\ref{fig:model_architecture} shows the overall view of SELENE. 

% \smallskip\noindent
\textbf{Challenges}.
Motivated by the empirical results in Section~\ref{sec:experimental_investigation}, we realise that a NE method for heterophilous networks should have the ability to distinguish nodes with different attributes or structural information. 
Each node's local structure can be flexibly defined by its $r$-ego network. 
We further summarise each node's relevant node attribute and structural information into an $r$-ego network and define the NE task as an $r$-ego network discrimination problem. 
Then, we address three main research challenges to solve this problem:
% \begin{enumerate}[leftmargin=\parindent,align=left,labelwidth=\parindent,labelsep=3pt]
% \begin{enumerate}
%     \item[\textbf{RC1}]
    (\textbf{RC1})
    How to leverage node attributes and network structure for NE?
    % \item[\textbf{RC2}]
    (\textbf{RC2})
    How to break the inherent homophily assumptions of traditional NE mechanisms?
    % \item[\textbf{RC3}]
    (\textbf{RC3})
    How to define an appropriate objective function to optimise the embedding learning process?
% \end{enumerate}

The following subsections discuss our solutions to address these three challenges and introduce the SELENE framework. 
In Section~\ref{sec:experiments}, we provide a comprehensive empirical evaluation on both synthetic and real data with varying homophily ratios to validate the effectiveness of SELENE under homophilous and heterophilous settings.
We also show that all components in our design are helpful in improving the quality of node embeddings.

\subsection{Dual-channel Feature Embedding (RC1)}
Motivated by the empirical observation in Section~\ref{sec:experimental_investigation} that AE (only based on raw node attributes) has the only stable performance across $h\in [0, 1)$ and graph structure-based NE approaches show priority when $h \to 1$. 
Therefore, to learn node embeddings that can discriminate $r$-ego networks of different nodes, we propose to deal with two important perspectives, i.e., node attributes and graph structure, separately. 
We first extract the $r$-ego network ($\mathcal{G}_{r}(v)$) of each node $v \in \mathcal{G}$.
% Specifically, we first define a subnetwork instance of a certain (ego) node $v$ to be its $r$-ego network $\mathcal{G}_{r}(v)$. 
For example, in Figure~\ref{fig:model_architecture}, $\mathcal{G}_{2}(v)$ represents a $2$-ego network instance of $\mathcal{G}$. 
% We note that given that $\mathcal{G}_{r}(v)$ captures the local structure of $v$, it is sufficient to compute a structural representation of $v$~\citep{QCDZYDWT20}.
% Then, we anonymise the sampled ego network $\mathcal{G}_{r}(v)$ by renaming its nodes to  $\{1, 2, \dots, |\mathcal{N}^{r}_{v}|\}$, in an arbitrary order. 
% Note that node order does not influence the representation quality because most GNNs are invariant to permutations of their inputs~\citep{BHBSZMT18}. 
% 
The empirical analysis of Section~\ref{sec:experimental_investigation} highlighted that node attributes and structural information play a major role in discriminating nodes over networks.
Therefore, we propose a dual-channel feature embedding pipeline to learn node representation from node attributes and network structure separately, as shown in Figure~\ref{fig:model_architecture}.
That said, we split $r$-ego network of each node $\mathcal{G}_{r}(v)=\{\mathcal{N}^{r}_v, \mathcal{E}^{r}_{v}, \mathbf{X}^{r}_{v} \}$ into ego node attribute $\mathbf{X}_{v}$ and network structure $\widetilde{\mathcal{G}}_{r}(v)=\{\mathcal{N}^{r}_v, \mathcal{E}^{r}_{v} \}$.
And in order to avoid dense ego networks, we adopt one widely adopted neighbour sampler~\citep{HYL17} to restrict the number of sampled nodes of each hop is no more than 15. 
Appendix~\ref{sec:appendix_experimental_setup_hyperparam_tune} will provide more description about the implementation. 
Such a design brings two main benefits: \textit{(i)} both important sources of information can be well utilised without interfering with each other, and \textit{(ii)} the inherent homophily assumptions of NE methods can be greatly alleviated, an issue we will address in the following subsection. 
We will empirically discuss the effectiveness of the dual channel feature embedding pipeline in Section~\ref{subsec:analysis}. 

\subsection{\textit{r}-ego Network Feature Extraction (RC2)}

% \smallskip\noindent
\textbf{Node attribute encoder module.}
As previously mentioned, learning effective node attribute representations is of great importance for NE.
% There are several alternative methods to learn representations for different types of data, including Autoencoder~\citep{HS06} and its derived variants~\citep{MMCS11,MSJG15,MRAVA16}.
% i.e., Convolutional Autoencoder~\citep{MMCS11}, LSTM Autoencoder~\citep{MRAVA16} and Adversarial Autoencoder~\citep{MSJG15}. 
In this paper, we employ the basic Autoencoder~\citep{HS06} to learn representations of raw node attributes, which can be replaced by more sophisticated encoders~\citep{MMCS11,MSJG15,MRAVA16} to obtain higher performance. 
We assume an $L$-layers Autoencoder ($f_{\theta}$), with the formulation of the $\ell$-th encoding layer being:
\begin{equation}
\label{eq:x_encoder}
    \mathbf{H}^{(\ell)}_e = \phi(\mathbf{W}^{(\ell)}_{e}\mathbf{H}^{(\ell-1)}_e+\mathbf{b}^{(\ell)}_{e})
\end{equation}
% $\mathbf{H}^{(\ell)}_e = \phi(\mathbf{W}^{(\ell)}_{e}\mathbf{H}^{(\ell-1)}_e+\mathbf{b}^{(\ell)}_{e})$
where $\phi$ is a non-linear activation function such as ReLU~\citep{NH10} or PReLU~\citep{HZRS15}.
$\mathbf{H}^{(\ell-1)}_e \in \mathbb{R}^{n \times d_{\ell-1}}$ is the hidden node attribute representations in layer $\ell-1$, with $d_{\ell-1}$ being the dimensionality of this layer's hidden representation. 
$\mathbf{W}^{(\ell)}_{e} \in \mathbb{R}^{d_{\ell-1} \times d_{\ell}}$ and $\mathbf{b}^{(\ell)}_{e} \in \mathbb{R}^{d_{\ell}}$ are trainable weight matrix and bias of the $\ell$-th layer in the encoder.
Node representations $\mathbf{H}_{v} = f_{\theta} (\mathbf{X}_{v}) = \mathbf{H}^{(L)}_{e}$ are obtained after successive application of $L$ encoding layers. 

% \smallskip\noindent
\textbf{Identity and network structure features.}
Despite the significant success of GNNs in a variety of network-related tasks, their representation power in network structural representation learning is limited~\citep{XHLJ19}. 
In order to obtain invariant node structural representations so that nodes with different ego network structures are assigned different representations, we employ the identity feature~\citep{YSYL21} and structural features~\citep{LWWL20}.
In this paper, we adopt variant shortest path distance (SPD) as node structural features ($\widetilde{\mathbf{X}}_{struc}$). 
After, we further inject the node identity features ($\widetilde{\mathbf{X}}_{id}$) as augment features, hence we have node feature matrix $\widetilde{\mathbf{X}} = \widetilde{\mathbf{X}}_{\it struc} + \widetilde{\mathbf{X}}_{\it id}$ for each ego-network. 

% \smallskip\noindent
\textbf{Network structure encoder module.}
Over the past few years, numerous GNNs have been proposed to learn node representations from network-structured data, including spectral GNNs
% ~\citep{DBV16,LMBB19,XSCQC19}
(i.e., ChebNet~\citep{DBV16}, CayletNet~\citep{LMBB19} and GWNN~\citep{XSCQC19}) 
and spatial GNNs
% ~\citep{HYL17,XHLJ19}.
(i.e., GraphSAGE~\citep{HYL17}, GAT~\citep{VCCRLB18}, GIN~\citep{XHLJ19}). 
For the sake of simplicity, we adopt a simple GNN variant, i.e., GCN~\citep{KW17}, as the building block of the network structure encoder ($f_{\delta}$). 
The $\ell$-th layer of a GCN for $v$'s $r$-ego network can be formally defined as:
\begin{equation}
\label{eq:gnn}
    \mathbf{U}^{(\ell)} = \sigma(\widehat{\mathbf{D}}^{-\frac{1}{2}}\widehat{\mathbf{A}}\widehat{\mathbf{D}}^{-\frac{1}{2}}\mathbf{U}^{(\ell-1)}\mathbf{W}^{(\ell)})
\end{equation}
with $\widehat{\mathbf{A}}=\mathbf{A}+\mathbf{I}$, where $\mathbf{I}$ is the identity matrix, and $\widehat{\mathbf{D}}$ is the diagonal node degree matrix of $\widehat{\mathbf{A}}$. 
$\textbf{U}^{(\ell-1)} \in \mathbb{R}^{n \times d_{\ell-1}}$ is the hidden representation of nodes in layer $\ell-1$, with $d_{\ell-1}$ being the dimensionality of this layer's representation, and $\mathbf{U}^{0}=\widetilde{\mathbf{X}}$. 
$\textbf{W}^{(\ell)} \in \mathbb{R}^{d_{\ell-1} \times d_{\ell}}$ is a trainable parameter matrix. 
$\sigma$ is a non-linear activation function such as ReLU or Sigmoid~\citep{HM95} function. 
Structural representations $\mathbf{U}_{v} = f_{\delta} (\widetilde{\mathcal{G}}_{r}(v)) = \mathbf{U}^{(L)}$ are obtained after successive applications of $L$ layers.

\subsection{Heterophily Adapted Self-Supervised Learning (RC3)}
\label{subsec:rc3}
The objective function plays a significant role in NE tasks. 
Several objective functions have been proposed for NE, such as network reconstruction loss~\citep{KW16}, distribution approximating loss~\citep{PAS14} and node distance approximating loss~\citep{HYL17}. 
Nevertheless, none of them is suitable for NE optimisation on heterophilous networks because of the homophily assumptions used to determine (dis)similar pairs.
In heterophilous networks, node distance on network alone does not determine (dis)similarity, i.e., connected nodes are not necessarily similar, and nodes far apart are not necessarily dissimilar. 
This removes the need for connected nodes to be close in the embedding space and for disconnected nodes to be far apart in the embedding space. 

Inspired by the latest success of negative-sample-free self-supervised learning (SSL), we adopt the Barlow-Twins (BT)~\citep{ZJMLD21} as our overall optimisation objective. 
Consequently, the invariance term makes the representation invariant to the distortions applied; the redundancy reduction term lets the representation units contain non-redundant information about the target sample. 
More details about the BT method can be found in Appendix~\ref{sec:appendix_more_about_bt_loss}. 
We discuss the feasibility of different objective functions in Section~\ref{subsec:analysis}. 

% \smallskip\noindent
\textbf{Distortion.}
To apply the BT method to optimise our framework, we have to first generate distorted samples for target $r$-ego network instances. 
Due to the dual-channel feature embedding pipeline, we introduce $f_{\it Aug}^{\mathbf{X}}$ and $f_{\it Aug}^{\mathcal{G}}$ to generate distorted node attribute and network structure instances, respectively.
Our data distortion method adopts a similar strategy as~\citet{YCSCWS20,BKC21} that randomly masks node attributes and edges with probability $p_x, p_e$. 
Formally, this distortion process can be represented as:
\begin{equation}
\begin{aligned}
    f_{\it Aug}^{\mathbf{X}}(\mathbf{X}_{v}, p_x) &= (\mathbf{X}^{(1)}_{v}, \mathbf{X}^{(2)}_{v}) \\
    f_{\it Aug}^{\mathcal{G}}(\widetilde{\mathcal{G}}_{r}(v), p_x, p_e) &= (\widetilde{\mathcal{G}}^{(1)}_{r}(v), \widetilde{\mathcal{G}}^{(2)}_{r}(v))
\end{aligned}
\end{equation}

% \smallskip\noindent
\textbf{Barlow-Twins loss function.}
Based on the two pairs of distorted instances, two pairs of representations ($\mathbf{U}^{(1)}_{v}$, $\mathbf{U}^{(2)}_{v}$) and ($\mathbf{H}^{(1)}_{v}$, $\mathbf{H}^{(2)}_{v}$) can be computed by applying $f_{\theta}$ and $f_{\delta}$, respectively. 
Then, BT method can be employed to evaluate the computed representations and guide the framework optimisation. 
Using representations ($\mathbf{H}^{(1)}$, $\mathbf{H}^{(2)}$) as example, the loss value is formally computed as:
\begin{equation}
\label{eq:barlow_twins_loss}
% \begin{aligned}
    \mathcal{L}_{BT}(\mathbf{H}^{(1)}, \mathbf{H}^{(2)}) = \sum^{|\mathcal{V}|}_{i}(1-\mathcal{C}_{ii})^{2}+\lambda\sum^{|\mathcal{V}|}_{i}\sum^{|\mathcal{V}|}_{j \neq i}\mathcal{C}^{2}_{ij}
    % \mathcal{L}_{BT}(\mathbf{H}^{(1)}, \mathbf{H}^{(2)}) &= \sum^{|\mathcal{V}|}_{i}(1-\mathcal{C}_{ii})^{2}+\lambda\sum^{|\mathcal{V}|}_{i}\sum^{|\mathcal{V}|}_{j \neq i}\mathcal{C}^{2}_{ij} \\
    % with \qquad \mathcal{C}_{ij} &= \frac{\sum_{b}\mathbf{H}_{b,i}^{(1)}\mathbf{H}_{b,j}^{(2)}}{\sqrt{\sum_{b}(\mathbf{H}_{b,j}^{(1)})^2}\sqrt{\sum_{b}(\mathbf{H}_{b,j}^{(2)})^2}}
% \end{aligned}
\end{equation}
where $\mathcal{C}_{ij} = \frac{\sum_{b}\mathbf{H}_{b,i}^{(1)}\mathbf{H}_{b,j}^{(2)}}{\sqrt{\sum_{b}(\mathbf{H}_{b,j}^{(1)})^2}\sqrt{\sum_{b}(\mathbf{H}_{b,j}^{(2)})^2}}$.
$\lambda>0$ defines the trade-off between the invariance and redundancy reduction terms, $b$ is the batch indexes, and $i$, $j$ index the vector dimension of the input representation vectors. 
We adopt the default settings as~\citet{ZJMLD21}. 

\textbf{Node attribute reconstruction loss function.}
In addition to the general objective function, $\mathcal{L}_{BT}$, we adopt another objective function, i.e., node attribute reconstruction loss $\mathcal{L}_{\it Rec}$, to optimise the node attribute encoder specifically.
Following the node attribute encoder (Eq.\ref{eq:x_encoder}), the decoder reconstructs input node attributes from the computed node representations $\mathbf{H}$.
Typically, a decoder has the same structure as the encoder by reversing the order of layers. 
Its $\ell$-th fully connected layer can be formally represented:
% \begin{equation}
% \label{eq:decoder}
%     \mathbf{H}^{(\ell)}_d = \phi(\mathbf{W}^{(\ell)}_{d}\mathbf{H}^{(\ell-1)}_{d}+\mathbf{b}^{(\ell)}_{d})
% \end{equation}
$\mathbf{H}^{(\ell)}_d = \phi(\mathbf{W}^{(\ell)}_{d}\mathbf{H}^{(\ell-1)}_{d}+\mathbf{b}^{(\ell)}_{d})$. 
% where $\mathbf{W}^{(\ell)}_{d}$ and $\mathbf{b}^{(\ell)}_{d}$ are trainable weight matrix and bias of $\ell$-th layer in the decoder, respectively. 
Reconstructed node attributes $\widehat{\mathbf{X}}=\mathbf{H}^{(L)}_d$ are obtained after successive applications of $L$ decoding layers. 
We optimise the autoencoder parameters by minimising the difference between raw node attributes $\mathbf{X}$ and reconstructed node attributes $\widehat{\mathbf{X}}$ with:
\begin{equation}
\label{eq:rec_loss}
    % \mathcal{L}_{\it Rec} = \frac{1}{2|\mathcal{V}|} \sum_{i=1}^{|\mathcal{V}|}||\mathbf{x}_i - \widehat{\mathbf{x}}_i||^{2}_{2}
    \mathcal{L}_{\it Rec}(\mathbf{X}, \widehat{\mathbf{X}}) = \frac{1}{2|\mathcal{V}|} ||\mathbf{X} - \widehat{\mathbf{X}}||^{2}_{F}
\end{equation}

Empowered with Barlow-Twins and node attribute reconstruction loss, we can optimise the framework's encoders under heterophilous settings. 
The node attribute encoder is optimised with $\mathcal{L}_{BT}$ (Eq.~\ref{eq:barlow_twins_loss}) and $\mathcal{L}_{\it Rec}$ (Eq.~\ref{eq:rec_loss}), and the network structure encoder is optimised with $\mathcal{L}_{BT}$ (Eq.~\ref{eq:barlow_twins_loss}).
The overall loss function is $\mathcal{L}=\mathcal{L}_{BT}(\mathbf{U}^{(1)}, \mathbf{U}^{(2)}) + \mathcal{L}_{BT}(\mathbf{H}^{(1)}, \mathbf{H}^{(2)}) + \mathcal{L}_{\it Rec}(\mathbf{X}, \widehat{\mathbf{X}})$. 
% Final node representations obtained by Eq.~\ref{eq:final_representation} contain information from node attributes and structural context, which will be used in downstream tasks. 

% \subsection{Final node representations}
% \label{subsec:final_node_representations}
% \smallskip\noindent
\textbf{Final node representations.}
Representations capturing node attributes ($\mathbf{H}$) and structural context ($\mathbf{U}$) are combined to obtain expressive and powerful representations as:
% \begin{equation}
% \label{eq:final_representation}
%     % \mathbf{Z} = (1-\alpha)\overline{\mathbf{H}} + \alpha\mathbf{U}
%     \mathbf{Z} = \textsc{COMBINE} (\mathbf{H}, \mathbf{U})
% \end{equation}
$\mathbf{Z} = \textsc{COMBINE} (\mathbf{H}, \mathbf{U})$
where \textsc{COMBINE}($\cdot$) can be any commonly used operation in GNNs~\citep{XHLJ19}, such as \textit{mean}, \textit{max}, \textit{sum} and \textit{concat}. 
We utilise \textit{concat} in all our experiments, allowing for an independent integration of representations learnt by the dual-channel architecture.
% The framework is comprehensively summarised in Algorithm~\ref{alg:framework_SELENE} in Appendix~\ref{sec:appendix_algorithm}. 

% \subsection{Model Scalability}
% \label{subsec:model_scalability}
% According to the design for SELENE, we can find that the network structure encoder module is categorised as a local network algorithm~\citep{T16}, which only involves local exploration of the network structure.
% On the other hand, the node attribute encoder module (Autoencoder) naturally support the mini-batch mechanism.
% Therefore, our design enables SELENE to scale to representation learning on large-scale networks and to be friendly to distributed computing settings~\citep{QCDZYDWT20}. 
% % The experimental results on large dataset further confirms this superiority (Section~\ref{subsec:experimental_results}). 

\subsection{Summary}
\label{subsec:summary}

\begin{algorithm}[!ht]
\SetAlgoLined
\KwIn{
	Network $\mathcal{G}=(\mathcal{V}, \mathcal{E}, \mathbf{X})$ \;
	Node attribute encoder $f_{\theta}$ and decoder $f_{\theta'}$ \;
	Network structure encoder (GNN) $f_{\delta}$ \;
	Node attribute distortion function $f^{\mathbf{x}}_{\it Aug}$ and network distortion function $f^{\mathcal{G}}_{\it Aug}$ \;
}
\KwOut{
	node representations $\mathbf{Z}$
}
Sample a set of $r$-ego networks $\{ \mathcal{G}_{r}(1), \mathcal{G}_{r}(2), \dots, \mathcal{G}_{r}(n) \}$ from $\mathcal{G}$ \;
Extract network structure of each $r$-ego network $\mathcal{G}_{r}(v)$ and enhance each with identity features ($\widetilde{\mathbf{X}}_{\it id}$) and network structure features ($\widetilde{\mathbf{X}}_{\it struc}$) to obtain $\widetilde{\mathcal{G}}_{r}(v) = \{\mathcal{N}^{r}_v, \mathcal{E}^{r}_{v}, \widetilde{\mathbf{X}} \}$ \; 
Generate distorted node attribute matrix instances $\mathbf{X}^{(1)}$ and $\mathbf{X}^{(2)}$ \;
Generate two distorted instances $(\widetilde{\mathcal{G}}_{r}^{(1)}(v), \widetilde{\mathcal{G}}_{r}^{(2)}(v))$ for each $r$-ego network $\widetilde{\mathcal{G}}_{r}(v)$ \;
\Repeat{Convergence}{
    Initialise loss $\mathcal{L}$ as zero \;
    \For{each node v $\in$ $\mathcal{V}$}{
        $\mathbf{H}_{v} = f_{\theta}(\mathbf{X}_{v})$, \hspace{10mm} $\widehat{\mathbf{X}}_{v} = f_{\theta'}(\mathbf{H}_{v})$ \;
        $\mathbf{H}^{(1)}_{v} = f_{\theta}(\mathbf{X}^{(1)}_{v})$, \hspace{6mm} $\mathbf{H}^{(2)}_{v} = f_{\theta}(\mathbf{X}^{(2)}_{v})$ \;
        $\mathbf{U}^{(1)}_{v} = f_{\delta}(\widetilde{\mathcal{G}}^{(1)}_{r}(v))$, \hspace{2mm} $\mathbf{U}^{(2)}_{v} = f_{\delta}(\widetilde{\mathcal{G}}^{(2)}_{r}(v))$ \;
        $\mathcal{L}=\mathcal{L}_{BT}(\mathbf{U}^{(1)}_{v}, \mathbf{U}^{(2)}_{v}) + \mathcal{L}_{BT}(\mathbf{H}^{(1)}_{v}, \mathbf{H}^{(2)}_{v}) + \mathcal{L}_{\it Rec}(\mathbf{X}_{v}, \widehat{\mathbf{X}}_{v})$ \;
    }
    Update $\theta$ and $\delta$ by descending the gradients $\nabla_{\theta, \delta} \mathcal{L}$ \;
}
$\mathbf{H} = f_{\theta}(\mathbf{X})$ \;
\For{each node v $\in$ $\mathcal{V}$}{
    $\mathbf{U}_{v} = f_{\delta}(\widetilde{\mathcal{G}}_{r}(v))$ \;
}
$\mathbf{Z} = \textsc{COMBINE} (\mathbf{H}, \mathbf{U})$ \;
\caption{\underline{SEL}f-sup\underline{E}rvised \underline{N}etwork \underline{E}mbedding (SELENE) Framework}
\label{alg:framework_SELENE}
% \vspace{-1mm}
\end{algorithm}

% We have presented the idea of SELENE and the design details of each component in Section~\ref{sec:approach}.
Here, we summarise SELENE in Algorithm~\ref{alg:framework_SELENE} to provide a general overview of our framework.
Given as input a graph $\mathcal{G}$ and node attribute encoder $f_{\theta}$ and decoder $f_{\theta'}$, network structure encoder $f_{\delta}$, node attribute distortion function $f^{\mathbf{X}}_{\it Aug}$, and network distortion function $f^{\mathcal{G}}_{\it Aug}$. 
Our algorithm is motivated by the empirical results in Section~\ref{sec:experimental_investigation}, which showed that a NE method for heterophilous networks should have the ability to distinguish nodes with different attributes or structural information. 
Therefore, we summarise each node's relevant node attribute and structural information into an $r$-ego network and define the NE task as an $r$-ego network discrimination problem. 
We sample a set of $r$-ego networks $\{ \mathcal{G}_{r}(1), \mathcal{G}_{r}(2), \dots, \mathcal{G}_{r}(n) \}$ from $\mathcal{G}$ (Line $1$).
Next, we extract network structure of each $r$-ego network $\mathcal{G}_{r}(v)$ and enhance each with identity features ($\widetilde{\mathbf{X}}_{\it id}$) and network structure features ($\widetilde{\mathbf{X}}_{\it struc}$) to obtain $\widetilde{\mathcal{G}}_{r}(v) = \{\mathcal{N}^{r}_v, \mathcal{E}^{r}_{v}, \widetilde{\mathbf{X}} \}$ (Line $2$).
Then, we generate distorted node attribute matrix instances $\mathbf{X}^{(1)}$ and $\mathbf{X}^{(2)}$ and two distorted instances $(\widetilde{\mathcal{G}}_{r}^{(1)}(v), \widetilde{\mathcal{G}}_{r}^{(2)}(v))$ for each $r$-ego network $\widetilde{\mathcal{G}}_{r}(v)$ (Line $3$-$4$). 
For each node $v \in \mathcal{G}$, we compute its node attribute embedding ($\mathbf{H}_v$) and distorted node attribute embeddings ($\mathbf{H}^{(1)}_v$, $\mathbf{H}^{(2)}_v$), and network structure embeddings ($\mathbf{U}^{(1)}_v$, $\mathbf{U}^{(2)}_v$) (Line $8$-$10$). 
Then, Barlow-Twins and network attribute reconstruction loss functions are employed to compute the final loss $\mathcal{L}$ (Line $11$). 
We update SELENE's parameters ($\theta$, $\delta$) by descending the gradient $\nabla_{\theta, \delta} \mathcal{L}$ (Line 13) until convergence. 
After the model is trained, we compute node attribute embedding $\mathbf{H}_v$ and network structure embedding $\mathbf{U}_v$ to obtain the final node representations $\mathbf{Z}$ by combining $\mathbf{H}$ and $\mathbf{U}$ (Line $15$-$19$). 

\textbf{Discussion.}
The description above indicates that the design of SELENE assumes nodes of the same class label share either similar node attributes or $r$-ego network structure. 
Under this assumption, we can learn node representations ($\mathbf{Z}$) to distinguish nodes of different class labels from the perspective of node attributes ($\mathbf{H}$) or $r$-ego network structure ($\mathbf{U}$). 
Nevertheless, if nodes of different class labels have similar node attributes and $r$-ego network structure, it is hard for SELENE to distinguish them. 
We address this limitation as a very interesting future work to explore.

%------------------------------------------------------------------------------

%------------------------------------------------------------------------------
\section{Evaluation}
\label{sec:experiments}
% We evaluate our proposed framework, SELENE, on benchmark real-world and synthetic datasets and compare with eleven competing methods over node clustering tasks.

\subsection{Real-world and Synthetic Datasets}
\label{subsec:datasets}

% \noindent
\textbf{Real-world datasets}.
We use a total of $12$ real-world datasets 
(Texas~\citep{PWCLY20}, Wisconsin~\citep{PWCLY20}, Actor~\citep{PWCLY20}, Chameleon~\citep{RAS21}, USA-Airports~\citep{RSF17}, Cornell~\citep{PWCLY20}, Europe-Airports~\citep{RSF17}, Brazil-Airports~\citep{RSF17}, Deezer-Europe~\citep{RS20}, Citeseer~\citep{KW17}, DBLP~\citep{FZMK20}, Pubmed~\citep{KW17}) 
in diverse domains (web-page, citation, co-author, flight transport and online user relation). 
All real-world datasets are available online\footnote{\url{https://pytorch-geometric.readthedocs.io/en/latest/modules/datasets.html}}.
Statistics information is summarised in Table~\ref{table:summary_real_datasets} of Appendix~\ref{sec:appendix_dataset_description}. 
% (see Appendix~\ref{sec:appendix_real_world_dataset_description}). 

% \smallskip\noindent
\textbf{Synthetic datasets}.
Moreover, we generate random synthetic networks with various homophily ratios $h$ by adopting a similar approach to~\citet{APKALHSG19,KO21}.
Specifically, each synthetic network has $10$ classes and $500$ nodes per class. 
Nodes are assigned random features sampled from 2D Gaussians, and each dataset has $10$ networks with $h \in [0, 0.1, 0.2, \dots, 0.9]$. 
% Here, we give detailed descriptions of the generation process.
The detailed data generation process can be found in Appendix~\ref{sec:appendix_dataset_description}. 

% \textit{Graph generation}.
% We generate synthetic graph $\mathcal{G}$ of $\vert\mathcal{V}\vert$ nodes with $\vert\mathcal{Y}\vert$ different class labels, and $\mathcal{G}$ has $\vert\mathcal{V}\vert/\vert\mathcal{Y}\vert$ nodes per class. 
% $\vert\mathcal{V}\vert$ and $\vert\mathcal{Y}\vert$ are two prescribed numbers to determine the size of $\mathcal{G}$. 
% A synthetic graph's homophily ratio $h$ is mainly controlled by $p_{in}$ and $p_{out}$, where $p_{in}$ means the possibility of existing an edge between two nodes with the same label and $p_{out}$ is the possibility of existing an edge between two nodes with different class labels. 
% Furthermore, the average degree of $\mathcal{G}$ is $d_{avg}=\vert\mathcal{V}\vert/\vert\mathcal{Y}\vert \cdot \delta$, where $\delta = p_{in} + (\vert\mathcal{Y}\vert-1) \cdot p_{out}$. 
% Following the described graph generation process, with given $\vert\mathcal{V}\vert$, $\mathcal{Y}$ and $d_{avg}$, we choose $p_{in}$ from $\{0.0001\delta, 0.1\delta, 0.2\delta, \dots, 0.9\delta \}$. 
% Note that the synthetic graph generation process requires both $p_{in}$ and $p_{out}$ are positive numbers, hence we use $p_{in}=0.0001\delta$ to estimate $h=0$. 

\subsection{Experimental Setup}
\label{subsec:experimental_setup}

% \noindent
% \textbf{Competing methods.}
We compare our framework SELENE
\footnote{Code and data are available at: \url{https://github.com/zhiqiangzhongddu/SELENE}} 
with $12$ competing NE methods in terms of the different challenging graph analysis tasks, including node clustering, node class prediction and link prediction. 
We adopt $2$ different competing NE methods without NNs, including node2vec (N2V)~\citep{GL16} and struc2vec (S2V)~\citep{RSF17}. 
We adopt $10$ additional competing NE methods using NNs, including AE~\citep{HS06}, GAE~\citep{KW16}, GraphSAGE (SAGE)~\citep{HYL17}, DGI~\citep{VFHLBH19}, SDCN~\citep{BWSZLC20}, GMI~\citep{PHLZRXH20}, GBT~\citep{BKC21}, H2GCN~\citep{ZYZHAK20}, FAGCN~\citep{BWSS21} and GPRGNN~\citep{CPLM21}. 
Note that GBT is an SSL approach that applies the Barlow-Twins~\citep{ZJMLD21} strategy to network-structured data. 
Albeit providing a new model training strategy, the basic GNN building blocks remain the same, hence still maintaining a homophily assumption. 
H2GCN, FAGCN and GPRGNN are state-of-the-art (SOTA) heterophilous GNN operators for supervised settings. 
Here, we train them using the same mechanism as GBT to adapt them to the unsupervised setting. 
We rename the unsupervised adaptations of H2GCN, FAGCN and GPRGNN as H2GCN$^*$, FAGCN$^*$ and GPRGNN$^*$, respectively.
See Appendix~\ref{sec:appendix_competing_method_description} and Appendix~\ref{sec:appendix_experimental_setup_hyperparam_tune} for details of all competing methods and implementation. 
% 
% \smallskip\noindent
% \textbf{Evaluation metrics.}
% We employ three node clustering evaluation metrics: accuracy (ACC), normalised mutual information (NMI) and average rand index (ARI).
% For each evaluation metric, a higher value means better node clustering performance. 

\subsection{Experimental Results}
\label{subsec:experimental_results}

\begin{table*}[!t]
\caption{Node clustering results on real-world datasets. 
The \textbf{bold} and \underline{underline} numbers represent the top-$2$ results.
OOM: out-of-memory.
}
% \vspace{-3mm}
% \small
\label{tab:clustering_results_real}
\resizebox{1\textwidth}{!}{
\begin{tabular}{c|c|cccccccccccc|c|c}
\toprule
Dataset & Metrics & AE & N2V & S2V & GAE & SAGE & SDCN & DGI & GMI & GBT & H2GCN$^*$ & FAGCN$^*$ & GPRGNN$^*$ & Ours & $\uparrow$ (\%) \\ 
\midrule
\multicolumn{15}{c}{Heterophilous datasets} \\
\midrule
\multirow{3}{*} {\specialcell{Texas \\ $h=0.11$}}
& ACC & 50.49 & 48.80 & 49.73 & 42.02 & 56.83 & 44.04 & 55.74 & 35.19 & 55.46 & \underline{58.80} & 57.92 & 57.50 & \textbf{65.23} & 10.94 \\
& NMI & 16.63 & 2.58 & 18.61 & 8.49 & 16.97 & 14.24 & 8.73 & 7.72 & 10.17 & 22.49 & \underline{23.35} & 22.83 & \textbf{25.40} & 8.78 	\\
& ARI & 14.6 & -1.62 & 20.97 & 10.83 & 23.50 & 10.65 & 8.25 & 2.96 & 12.10 & \underline{25.04} & 22.54 & 23.51 & \textbf{34.21} & 36.62 	\\
% \midrule
\hline
\multirow{3}{*} {\specialcell{Wisc. \\ $h=0.20$}}
& ACC & 58.61 & 41.39 & 43.03 & 37.81 & 46.29 & 38.25 & 44.58 & 36.97 & 48.01 & \underline{64.18} & 61.91 & 63.84 & \textbf{71.69} & 11.70 	\\
& NMI & \underline{30.92} & 4.23 & 11.23 & 9.19 & 10.16 & 8.46 & 10.72 & 11.68 & 7.55 & 29.64 & 27.35 & 29.54 & \textbf{39.51} & 27.78 	\\
& ARI & 28.53 & -0.48 & 11.50 & 5.2 & 6.06 & 3.67 & 10.31 & 3.74 & 3.85 & \underline{32.61} & 31.56 & 30.53 & \textbf{43.48} & 33.33 	\\
% \midrule
\hline
\multirow{3}{*} {\specialcell{Actor \\ $h=0.22$}}
& ACC & 24.19 & 25.02 & 22.49 & 23.45 & 23.08 & 23.67 & 24.26 & 26.18 & 24.68 & 25.55 & 25.61 & \underline{25.80} & \textbf{29.03} & 12.52 	\\
& NMI & 0.97 & 0.09 & 0.04 & 0.18 & 0.58 & 0.08 & 1.38 & 0.20 & 0.74 & \underline{3.23} & 3.22 & 3.21 & \textbf{4.72} & 46.13 	\\
& ARI & \underline{0.50} & 0.06 & -0.05 & -0.04 & 0.22 & -0.01 & 0.07 & 0.41 & -0.57 & 0.31 & 0.34 & 0.31 & \textbf{1.84} & 268.00 	\\
% \midrule
\hline
\multirow{3}{*} {\specialcell{Chamel. \\ $h=0.23$}}
& ACC & \underline{35.68} & 21.31 & 26.34 & 32.76 & 31.04 & 33.5 & 27.77 & 25.73 & 32.21 & 30.62 & 31.33 & 34.62 & \textbf{38.97} & 9.22 	\\
& NMI & 10.38 & 0.34 & 3.55 & 11.60 & 10.55 & 9.57 & 4.42 & 2.5 & 10.56 & 14.62 & \underline{14.71} & 10.31 & \textbf{20.63} & 40.24 \\
& ARI & 5.80 & 0.02 & 1.82 & 4.65 & 6.16 & 5.86 & 1.85 & 0.52 & \underline{7.01} & 4.78 & 5.16 & 5.01 & \textbf{15.94} & 127.39 	\\
% \midrule
\hline
\multirow{3}{*} {\specialcell{USA-Air. \\ $h=0.25$}}
& ACC & \underline{55.24} & 26.29 & 27.58 & 30.84 & 32.96 & 33.52 & 33.36 & 28.69 & 34.96 & 39.01 & 38.82 & 39.63 & \textbf{58.90} & 6.63	\\
& NMI & \underline{30.13} & 0.25 & 0.44 & 2.71 & 2.67 & 5.21 & 5.52 & 0.6 & 5.27 & 12.43 & 12.30 & 12.02 & \textbf{31.17} & 3.45 \\
& ARI & \underline{24.20} & -0.05 & 0.09 & 2.67 & 2.52 & 1.93 & 4.95 & 0.29 & 3.42 & 9.48 & 9.33 & 9.02 & \textbf{25.53} & 5.49 \\
% \midrule
\hline
\multirow{3}{*} {\specialcell{Cornell \\ $h=0.31$}}
& ACC & 52.19 & 50.98 & 32.68 & 43.72 & 44.7 & 36.94 & 44.1 & 33.55 & 52.19 & 54.97 & \underline{56.23} & 55.33 & \textbf{57.96} & 3.08 \\
& NMI & 17.08 & 5.84 & 1.54 & 5.11 & 4.33 & 6.6 & 5.79 & 5.26 & 5.94 & 17.05 & \underline{17.08} & 16.90 & \textbf{17.32} & 1.41 \\
& ARI & 17.41 & 0.18 & -2.20 & 6.51 & 5.64 & 3.38 & 4.87 & 3.05 & 0.63 & 19.50 & \underline{19.88} & 19.21 & \textbf{23.03} & 15.85 \\
% \midrule
\hline
\multirow{3}{*} {\specialcell{Eu.-Air. \\ $h=0.31$}}
& ACC & \underline{55.36} & 30.78 & 36.89 & 34.84 & 31.75 & 37.37 & 35.59 & 35.34 & 39.75 & 37.27 & 42.11 & 36.63 & \textbf{57.80} & 4.41 \\
& NMI & \underline{32.44} & 3.69 & 6.15 & 10.15 & 2.10 & 8.45 & 10.77 & 11.08 & 9.44 & 9.08 & 16.81 & 11.60 & \textbf{34.25} & 5.58 \\
& ARI & \underline{24.24} & 0.83 & 4.49 & 7.37 & 1.16 & 5.31 & 8.44 & 8.18 & 7.87 & 5.3 & 11.98 & 9.41 & \textbf{25.69} & 5.98 \\
% \midrule
\hline
\multirow{3}{*} {\specialcell{Bra.-Air. \\ $h=0.31$}}
& ACC & \underline{71.68} & 30.38 & 38.93 & 36.64 & 37.02 & 38.7 & 37.1 & 38.93 & 40.92 & 43.97 & 44.2 & 46.42 & \textbf{79.12} & 10.38 \\
& NMI &\underline{49.26} & 2.5 & 10.23 & 10.96 & 6.89 & 14.05 & 10.64 & 12.62 & 12.16 & 22.37 & 22.67 & 20.10 & \textbf{55.90} & 13.48 \\
& ARI &\underline{42.93} & -0.22 & 5.45 & 6.56 & 4.18 & 7.27 & 7.02 & 9.11 & 8.31 & 13.99 & 14.4 & 15.06 & \textbf{53.21} & 23.95 \\
% \midrule
\hline
\multirow{3}{*} {\specialcell{Deezer. \\ $h=0.53$}}
& ACC & 55.88 & 52.97 & OOM & 51.51 & 51.06 & 54.76 & 53.16 & OOM & OOM & 56.81 & \underline{56.81} & 53.22 & \textbf{59.94} & 5.51 \\
& NMI & \underline{0.28} & 0.0 & OOM & 0.13 & 0.16 & 0.17 & 0.05 & OOM & OOM & 0.27 & 0.27 & 0.01 & \textbf{0.34} & 21.43 \\
& ARI & 0.81 & 0.02 & OOM & 0.07 & -0.02 & 0.61 & -0.23 & OOM & OOM & \underline{1.22} & 0.82 & 0.77 & \textbf{1.33} & 9.02 \\
\midrule
\multicolumn{15}{c}{Homophilous datasets} \\
\midrule
\multirow{3}{*} {\specialcell{Citeseer \\ $h=0.74$}}
& ACC & 58.79 & 20.76 & 21.22 & 48.37 & 49.28 & \underline{59.86} & 58.94 & 59.04 & 57.21 & 47.33 & 47.42 & 45.83 & \textbf{60.02} & 0.27 \\
& NMI & 30.91 & 0.35 & 1.18 & 24.59 & 22.97 & 30.37 & \underline{32.6} & 32.11 & 31.9 & 20.48 & 20.18 & 20.58 & \textbf{32.74} & 0.31 \\
& ARI & 30.29 & -0.01 & 0.17 & 19.50 & 19.21 & 29.7 & 33.16 & 33.09 & \underline{33.17} & 18.03 & 17.93 & 18.11 & \textbf{33.46} & 0.87 \\
% \midrule
\hline
\multirow{3}{*} {\specialcell{DBLP \\ $h=0.80$}}
& ACC & 48.50 & 29.19 & 31.65 & 57.81 & 48.68 & 61.94 & 58.22 & 63.28 & \underline{73.10} & 41.87 & 41.25 & 43.96 & \textbf{75.74} & 3.61 \\
& NMI & 18.98 & 0.14 & 1.33 & 28.94 & 16.46 & 27.13 & 29.98 & 33.91 & \underline{42.21} & 11.04 & 10.60 & 11.27 & \textbf{44.17} & 4.64 \\
& ARI & 15.15 & -0.04 & 1.39 & 18.78 & 13.38 & 27.77 & 26.81 & 28.77 & \underline{42.57} & 4.67 & 4.32 & 4.92 & \textbf{46.35} & 8.88 \\
% \midrule
\hline
\multirow{3}{*} {\specialcell{Pubmed \\ $h=0.80$}}
& ACC & 65.34 & 39.32 & 37.39 & 42.08 & \underline{67.66} & 61.9 & 65.47 & OOM & OOM & 56.71 & 56.88 & 58.81 & \textbf{67.98} & 0.47 \\
& NMI & 26.89 & 0.02 & 0.07 & 1.28 & \underline{30.71} & 19.71 & 28.05 & OOM & OOM & 17.15 & 16.91 & 17.01 & \textbf{31.14} & 1.40 \\
& ARI & 25.98 & 0.09 & 0.06 & 0.15 & \underline{29.10} & 18.63 & 27.25 & OOM & OOM & 16.51 & 16.27 & 16.24 & \textbf{29.79} & 2.37 \\
\bottomrule
\end{tabular}
}
\vspace{-5mm}
\end{table*}

% \smallskip\noindent
% \textbf{Real-world datasets.}
\textbf{Node clustering.}
Node clustering results on homophilous and heterophilous \textit{real-world datasets} are summarised in Table~\ref{tab:clustering_results_real} where we see that SELENE is the best-performing method in all heterophilous datasets. 
In particular, compared to the best results of competing models, our framework achieves a significant improvement of up to $12.52\%$ on ACC, $46.13\%$ on NMI and $268\%$ on ARI. 
Such outstanding performance gain demonstrates that SELENE successfully integrates the important node attributes and network structure information into node representations. 
An interesting case is the comparison of SELENE, H2GCN$^*$, FAGCN$^*$ and GBT, given that H2GCN$^*$, FAGCN$^*$ and GBT also utilise the Barlow-Twins objective function to optimise a GNN model, and the major difference between H2GCN$^*$, FAGCN$^*$, GBT and SELENE is our designs to address \textbf{RC1} and \textbf{RC2}.
SELENE has a superior performance in all heterophilous datasets, indicating the effectiveness of designs and showing that simply porting supervised models to unsupervised scenarios is not appropriate. 
Results of node clustering on homophilous networks, contained in Table~\ref{tab:clustering_results_real}, show that SELENE achieves new SOTA performance.
This demonstrates SELENE's suitability for homophilous networks and proves the flexibility of our designs.
% 
% \smallskip\noindent
% \textbf{Synthetic networks.}
For the \textit{Synthetic networks},
we present SELENE vs SDCN (SOTA node clustering model) vs GBT (SOTA graph contrastive learning model) vs FAGCN$^*$ (SOTA heterophilous GNN model) clustering accuracy in Figure~\ref{fig:results}-(a). 
SELENE achieves the best performance on the synthetic datasets, which shows SELENE adapts well to homophily/heterophily scenarios with(out) contextually raw node attributes. 
Moreover, FAGCN$^*$ performs worse on $10$ synthetic networks, which indicates that the heterophilous GNN models designed for supervised settings do not adapt well to unsupervised settings because they need supervision information to train the more complex aggregation mechanism.
In addition, we observe that SELENE has no significant improvement with $h \leq 0.4$ because the network structure encoder's expressive power on extreme homophily is limited.
Overall, SELENE still works significantly better for homophilous graphs, but deteriorates more gracefully compared to prior works while $h \to 0$.

% In Section~\ref{subsec:experimental_results}, we presented the node clustering results of various models. 
\textbf{Node classification and link prediction.}
Meanwhile, we acknowledge that some works adopt a different experiment pipeline that utilises node labels to train a classifier, after obtaining unsupervised node representations, to predict labels of test nodes~\citep{VFHLBH19,PHLZRXH20,BKC21}.
% We argue that this is not a precise setting because if some node labels are available to train a classifier, we can directly utilise them to train the NE model (the typical supervised setting). 
% However, hereby we still provide results of a number of real-world datasets that follow this setting to show the generality of SELENE.
Hence, we provide results of a number of real-world datasets that follow this setting to show the generality of SELENE.
% More discussion about experimental settings refers to Appendix~\ref{sec:appendix_more_experimental_results_on_different_settings}.
% 
Results in Table~\ref{tab:classification_results_real} show that SELENE achieves the best classification performance on all datasets with homophily and heterophily. 
Moreover, we also report another major graph analysis task, i.e., link prediction, performances of a number of real-world datasets in Table~\ref{tab:link_prediction_results_real}.
From the experimental results, we can find that \textit{(i)} the homophily ratio has a significant influence on the link prediction task, and \textit{(ii)} SELENE performs priority on this task as well. 
For more experimental settings and experimental results discussion, refer to Appendix~\ref{sec:appendix_more_experimental_results_on_different_settings}.

\begin{table*}[!ht]
\caption{Node classification results on real-world datasets. 
The \textbf{bold} and \underline{underline} numbers represent the top-$2$ results.
}
% \vspace{-3mm}
% \small
\label{tab:classification_results_real}
\resizebox{1\textwidth}{!}{
\begin{tabular}{c|c|cccccccccccc|c|c}
\toprule
Dataset & F1-score & AE & N2V & S2V & GAE & SAGE & SDCN & DGI & GMI & GBT & H2GCN$^*$ & FAGCN$^*$ & GPRGNN$^*$ & Ours & $\uparrow$ (\%)  \\ 
\midrule
\multicolumn{15}{c}{Heterophilous datasets} \\
\midrule
\multirow{2}{*} {Texas}
& Micro & 0.598 & 0.509 & 0.592 & 0.5 & 0.563 & 0.552 & 0.541 & 0.505 & 0.461 & 0.523 & \underline{0.607} & 0.561 & \textbf{0.643} & 5.93 \\
& Macro  & 0.284 & 0.165 & \underline{0.335} & 0.253 & 0.301 & 0.153 & 0.266 & 0.295 & 0.253 & 0.186 & 0.265 & 0.262 & \textbf{0.381} & 13.73 	\\
\hline
\multirow{2}{*} {Actor}
& Micro & \underline{0.284} & 0.240 & 0.227 & 0.254 & 0.277 & 0.252 & 0.272 & 0.272 & 0.245 & 0.244 & 0.281 & \underline{0.284} & \textbf{0.341} & 20.07 	\\
& Macro & 0.194 & 0.178 & 0.190 & 0.164 & 0.204 & 0.125 & \underline{0.244} & 0.195 & 0.224 & 0.138 & 0.166 & 0.192 & \textbf{0.274} & 12.30 	\\
\hline
\multirow{2}{*} {USA-Air.}
& Micro & \underline{0.541} & 0.239 & 0.254 & 0.309 & 0.311 & 0.302 & 0.338 & 0.253 & 0.35 & 0.535 & 0.527 & 0.453 & \textbf{0.565} & 4.44	\\
& Macro & 0.476 & 0.235 & 0.253 & 0.236 & 0.299 & 0.288 & 0.260 & 0.221 & 0.288 & \underline{0.483} & 0.454 & 0.370 & \textbf{0.532} & 10.14 \\
\hline
\multirow{2}{*} {Bra.-Air.}
& Micro & \underline{0.611} & 0.247 & 0.301 & 0.331 & 0.312 & 0.304 & 0.321 & 0.329 & 0.371 & 0.465 & 0.507 & 0.467 & \textbf{0.715} & 17.02 \\
& Macro & \underline{0.538} & 0.203 & 0.284 & 0.244 & 0.295 & 0.281 & 0.244 & 0.249 & 0.319 & 0.415 & 0.420 & 0.426 & \textbf{0.692} & 28.62 \\
\midrule
\multicolumn{15}{c}{Homophilous datasets} \\
\midrule
\multirow{2}{*} {DBLP}
& Micro & 0.746 & 0.269 & 0.334 & 0.768 & 0.732 & 0.638 & 0.792 & \underline{0.811} & 0.744 & 0.413 & 0.560 & 0.643 & \textbf{0.813} & 0.25 \\
& Macro  & 0.737 & 0.227 & 0.305 & 0.758 & 0.721 & 0.633 & 0.784 & \underline{0.804} & 0.736 & 0.371 & 0.536 & 0.585 & \textbf{0.810} & 0.75  \\
\bottomrule
\end{tabular}
}
\vspace{-3mm}
\end{table*}

\begin{table*}[!ht]
\caption{Link prediction results on real-world datasets. 
The \textbf{bold} and \underline{underline} numbers represent the top-$2$ results.
}
% \vspace{-3mm}
% \small
\label{tab:link_prediction_results_real}
\resizebox{1\textwidth}{!}{
\begin{tabular}{c|c|cccccccccccc|c}
\toprule
Dataset & Metric & AE & N2V & S2V & GAE & SAGE & SDCN & DGI & GMI & GBT & H2GCN$^*$ & FAGCN$^*$ & GPRGNN$^*$ & Ours \\ 
\midrule
\multicolumn{14}{c}{Heterophilous datasets} \\
\midrule
Texas
& ROC-AUC  & 0.529 & 0.549 & 0.571 & 0.634 & 0.559 & 0.640 & \textbf{0.787} & 0.777 & 0.783 & 0.451 & 0.673 & 0.782 & \underline{0.784}	\\
\hline
Actor & ROC-AUC  & 0.501 & 0.673 & 0.652 & 0.668 & 0.634 & 0.552 & 0.626 & 0.609 & 0.598 & 0.679 & 0.613 & \textbf{0.796} & \underline{0.725}	\\
\hline
USA-Air.
& ROC-AUC  & 0.518 & 0.714 & 0.685 & 0.842 & 0.595 & 0.528 & 0.857 & 0.854 & 0.867 & 0.667 & 0.642 & \textbf{0.925} & \underline{0.871} 	\\
\hline
Bra.-Air.
& ROC-AUC  & 0.665 & 0.592 & 0.637 & 0.787 & 0.615 & 0.677 & \underline{0.855} & 0.791 & 0.786 & 0.702 & 0.696 & 0.838 & \textbf{0.881} 	\\
\midrule
\multicolumn{14}{c}{Homophilous datasets} \\
\midrule
DBLP
& ROC-AUC  & 0.872 & 0.758 & 0.717 & 0.870 & 0.841 & 0.628 & \textbf{0.939} & 0.884 & 0.892 & 0.736 & 0.764 & 0.890 & \underline{0.908} 	\\
\bottomrule
\end{tabular}
}
\vspace{-3mm}
\end{table*}

\subsection{Analysis}
\label{subsec:analysis}

% \smallskip\noindent
\textbf{Model scalability.}
The network structure encoder module is categorised as a local network algorithm~\citep{T16}, which only involves local exploration of the network structure.
On the other hand, the node attribute encoder module naturally supports the mini-batch mechanism.
Therefore, our design enables SELENE to scale to representation learning on large-scale networks and to be friendly to distributed computing settings~\citep{QCDZYDWT20}. 
Table~\ref{tab:clustering_results_real} illustrates that three competing methods, i.e., struc2vec, GMI and GBT, have out-of-memory issues on large datasets, i.e., Pubmed and Deezer., an issue which did not arise with SELENE.
This shows SELENE's advantage in handling large-scale networks due to its local network algorithm characteristic. 
% Note that we only use one GPU in experiments, and such an advantage would be more evident in multi-GPU computing scenarios.

\begin{figure*}[!t]
\centering
\includegraphics[width=\linewidth]{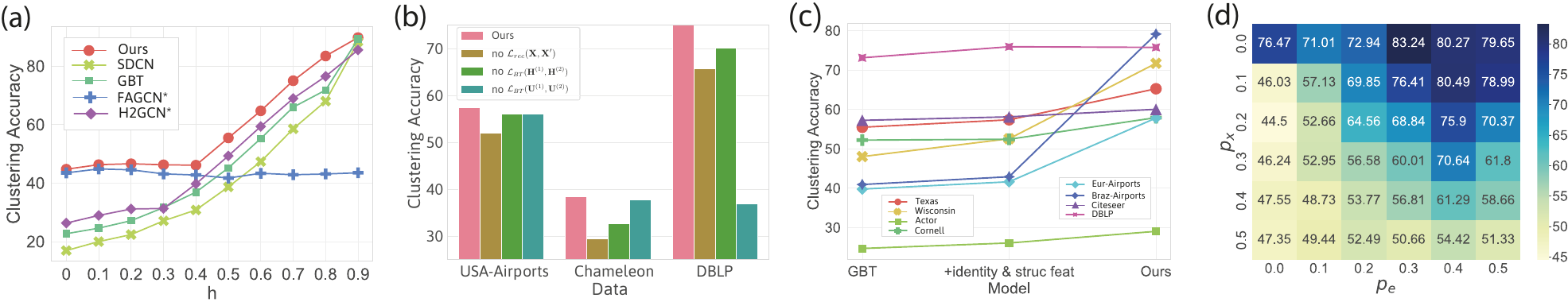}
\caption{(a) Clustering accuracy comparison on synthetic networks. (b) Loss function ablation on real-world networks. (c) Framework component exploration on real-world networks. (d) Hyperparameter influences exploration on synthetic-0.8.
}
\label{fig:results}
\vspace{-5mm}
\end{figure*}

\textbf{Effectiveness of loss function.}
The objective loss function of SELENE contains three components, and we thus sought to test the effectiveness of each component. 
In particular, we ablate each component and evaluate the obtained node representations on two heterophilous datasets (USA-Air., Chamel.) and one homophilous dataset (DBLP). 
Results shown in Figure~\ref{fig:results}-(b) indicate that the ablation of any component decreases the model's performance. 
Specifically, the ablation of $\mathcal{L}_{\it Rec}(\mathbf{X}, \widehat{\mathbf{X}})$ causes steeper performance degradation in heterophilous datasets, and the ablation of $\mathcal{L}_{BT}(\mathbf{U}^{(1)}, \mathbf{U}^{(2)})$ causes steeper performance degradation in homophilous datasets, which indicates the importance of node attributes and network structure information for heterophilous and homophilous networks, respectively (consistent with observations in Section~\ref{sec:experimental_investigation}).

% \smallskip\noindent
\textbf{Effectiveness of dual-channel feature embedding pipeline.}
SELENE contains a novel dual-channel features embedding pipeline to integrate node attributes and network structure information, thus, we conduct an ablation study to explore the effectiveness of this pipeline.
We first remove the pipeline and only use the Barlow-Twins loss function to train a vanilla GCN encoding module (such a structure is the same as GBT, hence we remark it as GBT). 
Next, we add the network structure channel, which includes $r$-ego network extraction, anonymisation and distortion.
Lastly, we add the node attribute channel to form the complete SELENE framework.
Experimental results are shown in Figure~\ref{fig:results}-(c). 
Overall, we observe that the design of each channel is useful for learning better representation, with the node attribute channel playing a major role in the embedding of heterophilous networks.
Note that adding the node attribute channel slightly decreases the clustering accuracy for the homophilous dataset, i.e., DBLP, but it is still competitive. 

% \smallskip\noindent
\textbf{Influence of $p_{x}$ and $p_e$.}
We present SELENE's clustering accuracy on synthetic-$0.8$ with different $p_{x}$ and $p_e$ in Figure~\ref{fig:results}-(d).
The figure indicates that hyperparameters of distortion methods significantly influence representation quality.

% \textbf{Additional experiments.}
% In Section~\ref{subsec:experimental_results}, we presented the node clustering results of various models. 
% Meanwhile, we acknowledge that some works adopt a different experiment pipeline that utilises node labels to train a classifier, after obtaining unsupervised node representations, to predict labels of test nodes~\citep{VFHLBH19,PHLZRXH20,BKC21}.
% We argue that this is not a precise setting because if some node labels are available to train a classifier, we can directly utilise them to train the NE model (the typical supervised setting). 
% However, hereby we still provide results of a number of real-world datasets that follow this setting to show the generality of SELENE.
% % More discussion about experimental settings refers to Appendix~\ref{sec:appendix_more_experimental_results_on_different_settings}.
% % 
% Results in Table~\ref{tab:classification_results_real} show that SELENE achieves the best classification performance on all datasets with homophily and heterophily. 
% % 
% Moreover, we also report another major graph analysis task, i.e., link prediction, performances of a number of real-world datasets in Table~\ref{tab:link_prediction_results_real}.
% From the experimental results, we can find that \textit{(i)} the homophily ratio has a significant influence on the link prediction task, and \textit{(ii)} SELENE performs priority on this task as well. 
% For more experimental settings and experimental results discussion, refer to Appendix~\ref{sec:appendix_more_experimental_results_on_different_settings}. 

% \smallskip\noindent
\textbf{Effectiveness of heterophily adopted self-supervised learning.}
We intuitively discussed the importance of selecting a proper objective function for heterophily NE tasks in Section~\ref{subsec:rc3} and explained the reason for selecting reconstruction loss (Eq.~\ref{eq:rec_loss}) and BT loss (Eq.~\ref{eq:barlow_twins_loss}). 
Experimental results demonstrate the effectiveness of heterophily adopted self-supervised learning objective functions. 
For instance, SELENE shows priority compared with the network reconstruction loss-based model (GAE), distribution approximating loss-based model (N2V), node distance approximating loss-based model (SAGE), negative-sampling-based model (DGI) and mutual information-based model (GMI).

%------------------------------------------------------------------------------

%------------------------------------------------------------------------------
\section{Conclusion and Future Directions}
\label{sec:conclusion}
In this paper, we focused on the unsupervised network embedding task with challenging heterophily settings and tackled two main research questions.
First, we showed through an empirical investigation that the performance of existing embedding methods that utilise network structure decreases significantly with the decrease of network homophily ratio.
Second, to address the identified limitations, we proposed SELENE, which effectively fuses node attributes and network structure information without additional supervision.
Comprehensive experiments demonstrated the significant performance of SELENE, and additional ablation analysis confirms the effectiveness of components of SELENE on real-world and synthetic networks.
As future work, \textit{(i)} we propose to explore the \textit{Information Bottleneck}~\citep{TZ15,WRLL20} of network embedding to theoretically define the optimal representation of an arbitrary network 
% and optimally balance expressiveness and robustness against potential external attacks
~\citep{ZAG18};
\textit{(ii)} designing a more powerful unsupervised network structure encoder for extreme heterophilous networks is also a promising future work. 

%------------------------------------------------------------------------------

\subsubsection*{Acknowledgments}
This work is supported by the Luxembourg National Research Fund through grant PRIDE15/10621687/SPsquared. The authors (G.G.) were supported by the ERC-Consolidator Grant No. 724228 (LEMAN).

\bibliography{full_format_references}
\bibliographystyle{tmlr}

\clearpage
%------------------------------------------------------------------------------
\appendix

\section{Dataset Description and Evaluation Metrics}
\label{sec:appendix_dataset_description}

In our experiments, we use the following real-world datasets with varying homophily ratios $h$.
Their statistics are provided in Table~\ref{table:summary_real_datasets}. 

\begin{table}[!ht]
\caption{Statistics of real-world datasets.
$|\mathcal{V}|$: number of nodes;
$|\mathcal{E}|$: number of edges;
$\pi$: dimensionality of nodes features;
OSF: nodes only have structure related features;
$d_{\it avg}$: average degree;
$|\mathcal{Y}|$: number of possible class labels;
$h$: homophily ratio;
}
\label{table:summary_real_datasets}
\centering
% \footnotesize
\small
% \resizebox{1.\linewidth}{!}{
\begin{tabular}{l|r|r|r|r|r|r|r}
\toprule 
\textbf{Dataset}     & \textbf{$|\mathcal{V}|$}   & \textbf{$|\mathcal{E}|$}   & \textbf{$\pi$} & OSF & \textbf{$|\mathcal{Y}|$} & \textbf{$d_{\it avg}$} & \textbf{$h$} \\
\midrule
Texas       & 183       & 325       & 1,703         & False & 5         & 1.8    & 0.108 \\
Wisconsin   & 251       & 515       & 1,703         & False & 5         & 2.1    & 0.196 \\
Actor       & 7,600     & 30,019    & 932           & False & 5         & 3.9    & 0.219 \\
Chameleon   & 2,277     & 31,421    & 2,325          & False & 5         & 27.6   & 0.233 \\
USA-Airports & 1,190    & 13,599    & 1             & True  & 4         & 22.9   & 0.251 \\
Cornell     & 183       & 298       & 1,703         & False & 5         & 1.6    & 0.305 \\
Europe-Airports & 399   & 11,988     & 1             & True  & 4         & 30.1   & 0.309 \\
Brazil-Airports & 131   & 2,077      & 1             & True  & 4         & 16.4   & 0.311 \\
Deezer-Europe& 28,281   & 185,504   & 31,241        & False & 2         & 6.6    & 0.525 \\
\midrule
Citeseer    & 3,327     & 4,552     & 3,703         & False & 6         & 2.7    & 0.736 \\
DBLP        & 4,057     & 3,528     & 334           & False & 4         & 1.7    & 0.799 \\
Pubmed      & 19,717    & 88,648    & 500           & False & 3         & 4.5    & 0.802 \\
% \hline
\bottomrule
\end{tabular}
% }
% \vspace{-5mm}
\end{table}

\begin{itemize}[leftmargin=*]\itemsep0em 
    \item \textbf{Texas}, \textbf{Wisconsin} and \textbf{Cornell}~\citep{ZYZHAK20} are networks that represent links between web pages of the corresponding universities, originally collected by the CMU WebKB project. 
    Nodes of these networks are web pages, which are classified into $5$ categories: course, faculty, student, project and staff. 
    \item \textbf{Chameleon}~\citep{ZYZHAK20} is a subgraph of web pages in Wikipedia discussing the corresponding topics
    \item \textbf{Actor}~\citep{ZYZHAK20} is the actor-only reduced subgraph of the film-director-actor-writer network~\citep{TJWY09}. 
    Each node corresponds to an actor, node features are keywords of the actor's Wikipedia page, and edges mean whether two actors denote co-occurrence on the same page. 
    Nodes are classified into five categories in terms of words from actors' Wikipedia.
    \item \textbf{USA-Airports}, \textbf{Europe-Airports}, \textbf{Brazil-Airports} are three air traffic networks collected from government websites throughout the year 2016 and were used to evaluate algorithms to learn structure representations of nodes. 
    Nodes represent airports, and edges indicate whether there are commercial flights between them. 
    The goal is to infer the level of an airport using the connectivity pattern solely.
    \item \textbf{Deezer-Europe}~\citep{ZLPZJY22} is a social network of users on Deezer from European countries, where edges represent mutual follower relationships.
    The node features are based on artists liked by each user.
    Nodes are labelled with reported gender.
    \item \textbf{Citeseer} and \textbf{Pubmed} are papers citation networks that were originally introduced in~\citet{KW17}, which contains sparse bag-of-words feature vectors for each document and a list of citation links between these documents.
    \item \textbf{DBLP}~\citep{BWSZLC20} is an author network from the DBLP dataset.
    There is an edge between the two authors if they have a co-author relationship in the dataset.
    And the authors are divided into four research areas: database, data mining, artificial intelligence, and computer vision.
    We label each author's research area depending on the conferences they submitted. Author features are the elements of a bag of words represented by keywords.
\end{itemize}

\textit{Graph generation}.
We generate synthetic graph $\mathcal{G}$ of $\vert\mathcal{V}\vert$ nodes with $\vert\mathcal{Y}\vert$ different class labels, and $\mathcal{G}$ has $\vert\mathcal{V}\vert/\vert\mathcal{Y}\vert$ nodes per class. 
$\vert\mathcal{V}\vert$ and $\vert\mathcal{Y}\vert$ are two prescribed numbers to determine the size of $\mathcal{G}$. 
A synthetic graph's homophily ratio $h$ is mainly controlled by $p_{in}$ and $p_{out}$, where $p_{in}$ means the possibility of existing an edge between two nodes with the same label and $p_{out}$ is the possibility of existing an edge between two nodes with different class labels. 
Furthermore, the average degree of $\mathcal{G}$ is $d_{avg}=\vert\mathcal{V}\vert/\vert\mathcal{Y}\vert \cdot \delta$, where $\delta = p_{in} + (\vert\mathcal{Y}\vert-1) \cdot p_{out}$. 
Following the described graph generation process, with given $\vert\mathcal{V}\vert$, $\mathcal{Y}$ and $d_{avg}$, we choose $p_{in}$ from $\{0.0001\delta, 0.1\delta, 0.2\delta, \dots, 0.9\delta \}$. 
Note that the synthetic graph generation process requires both $p_{in}$ and $p_{out}$ are positive numbers, hence we use $p_{in}=0.0001\delta$ to estimate $h=0$. 

% \smallskip\noindent
\textbf{Evaluation metrics.}
We employ three node clustering evaluation metrics: accuracy (ACC), normalised mutual information (NMI) and average rand index (ARI).
For each evaluation metric, a higher value means better node clustering performance.

\section{Competing Methods Description}
\label{sec:appendix_competing_method_description}
% We adopt $8$ competing NE methods:
\begin{itemize}[leftmargin=*]\itemsep0em 
\item
% \smallskip\noindent
\textbf{AE}~\citep{HS06}: A classical deep representation learning method relies on raw node attributes with an encoder-decoder pipeline.

\item
% \smallskip\noindent
\textbf{node2vec}~\citep{GL16}: A random-walk based shallow node representation learning method with two hyper-parameter to adjust the random walk.

\item
% \smallskip\noindent
\textbf{struc2vec}~\citep{RSF17}: A shallow node representation learning method that learns node representations by identifying node's related local structure and estimating their relationship to other nodes. 
It relies on network structure information.
 
\item
% \smallskip\noindent
\textbf{GAE}~\citep{KW16}: Deep node representation learning methods using GCN to learn node representations by reconstructing the network structure. 

\item
% \smallskip\noindent
\textbf{GraphSAGE}~\citep{HYL17}: A deep node representation learning method with a specifically designed loss function that makes nodes far apart have different representations. 
It relies on both node attributes and network structure.

\item 
% \smallskip\noindent
\textbf{DGI}~\citep{VFHLBH19}: A self-supervised learning model by migrating infomax techniques from computer vision domain to networks. 

\item 
% \smallskip\noindent
\textbf{GMI}~\citep{PHLZRXH20}: It generalises the idea of conventional mutual information computations from vector space to
the graph domain where measuring mutual information from two aspects of node features and topological structure is indispensable. 

\item
% \smallskip\noindent
\textbf{SDCN}~\citep{BWSZLC20}: Deep learning node representations methods specifically for node clustering task. 
An AE component relies on raw node attributes, and another GCN component merges structure and node attributes together and trains with a dual self-supervised module.
% Note that we report the best performance with/without pre-trained AE component for fair comparison. 
 
\item
% \smallskip\noindent
\textbf{GBT}~\citep{BKC21}: A self-supervised learning approach that migrates the Barlow-Twins approach to network structure data. 
However, they don't modify the model structure of GNNs but just provide a new mode training approach, hence still maintaining the homophily assumption. 

\item
% \smallskip\noindent
\textbf{H2GCN}~\citep{ZYZHAK20}: A heterophilous GNN operator that separately aggregates information from ego node, neighbouring nodes and high-order neighbouring nodes under supervised settings. 

\item
% \smallskip\noindent
\textbf{FAGCN}~\citep{BWSS21}: Different from conventional GNN operators that only have a low-frequency filter, they propose another high-frequency filter to collaboratively work with low-frequency filter to work on networks with homophily and heterophily, under supervised settings. 

\item
\textbf{GPRGNN}~\citep{CPLM21}: They introduce Generalised PageRank (GPR) GNN architecture to adaptively learn GPR weights to jointly optimise node feature and topological information extraction. 

\end{itemize}

\section{Experimental Setup \& Hyperparameter Tuning}
\label{sec:appendix_experimental_setup_hyperparam_tune}

% \smallskip\noindent
% \textbf{Model implementation.}
For network embedding methods without neural networks, we set the embedding dimension to $128$, the number of random walks of each node to $10$ and the walk length to $80$. 
For node2vec, we additionally select $p$, $q$ over $\{0.25, 0.5, 1, 2\}$ with best clustering performance. 
We utilise the integrated implementations from GraphEmbedding\footnote{\url{https://github.com/shenweichen/GraphEmbedding}}. 

We train network embedding methods with neural networks, including AE, GAE, VGAE and SDCN, with the same settings as~\citet{BWSZLC20}. 
Specifically, we train the models end-to-end using all nodes and edges with $30$ epochs and a learning rate of \num{e-3}.
For AE, we set the representation dimensions to $\{\pi-500-500-200-10 \}$, where $\pi$ is the dimensionality of raw node attributes. 
For GNN-related methods, including GAE, VGAE, GraphSAGE, DGI, GBT, GMI, FAGCN$^*$, H2GCN$^*$ and SELENE, we set their representation dimensions to $\{\pi-256-16 \}$. 
For GPRGNN$^*$, we adopt its default dimension setting of official implementation. 
The representation dimensions of SDCN's GCN module are the same as AE~\citep{BWSZLC20} proposes to pretrain the SDCN's AE component to boost its performance, thus, we report the best performance of SDCN with/without pre-trained AE.
For DGI, GMI, GBT and SELENE, we assign the same GCN~\citep{KW17} encoder and follow the optimisation protocol as~\citet{BKC21}. 
We set $r=3$ for ego network extraction following~\citet{LWWL20} and the batch size = $512$. 
 
For GAE, VGAE, GraphSAGE, DGI and FAGCN we use the implementation from Pytorch-Geometric\footnote{\url{https://pytorch-geometric.readthedocs.io/en/latest/}}; for AE \& SDCN and GMI, GBT and H2GCN, we use the implementation from the published code of SDCN\footnote{\url{https://github.com/bdy9527/SDCN}}, GMI\footnote{\url{https://github.com/zpeng27/GMI}}, GBT\footnote{\url{https://github.com/pbielak/graph-barlow-twins}} H2GCN\footnote{\url{https://github.com/GemsLab/H2GCN}} and GPRGNN\footnote{\url{https://github.com/jianhao2016/GPRGNN}}, respectively, 

Note that, all experiments are conducted on a single Tesla V100 GPU. 
For all NN-based methods, we initialise them $10$ times with random seeds and select the best solution to follow the similar setting as~\citet{BWSZLC20}. 
After obtaining node representations with each model, we feed the learned representations into a \textit{K}-means clustering model~\citep{HW79} to get the final clustering prediction. 
For the \textit{K}-means algorithm~\citep{HW79}), we follow the setting of \citet{BWSZLC20} that adopt the default Sklearn implementation\footnote{\url{https://scikit-learn.org/stable/modules/generated/sklearn.cluster.KMeans.html}}. 
In addition, the random seed of different parts of our pipeline is related to experimental performance, thus we design a $seed$ module in our code to set up fixed random seed for the relevant libraries, such as Numpy, Pytorch and Random. 
The final clustering section is thus repeated $10$ times with a set of seeds, and we report the mean performance. 

\textbf{$r$-ego network resolution.}
The introduced $r$-ego network sampling approach (Section~\ref{sec:approach}) is indispensable for the dual-channel feature embedding pipeline.
However, in experiments on relatively large dense dataset (i.e., datasets with high $|\mathcal{V}|$ and $d_{avg}$), we observe that an entire 3-ego network can sometimes fail to into the GPU memory with the same batch size setting, and will lead to heavily training process. 
Therefore, we practically consider one of the most widely adopted samplers-neighbours sampler~\citep{HYL17}, where the number of sampled neighbours are limited to no more than 15 at each hop.

\section{Additional Experimental Results and Discussion}
\label{sec:appendix_more_experimental_results_on_different_settings}

\begin{figure*}[!ht]
\centering
\includegraphics[width=0.6\linewidth]{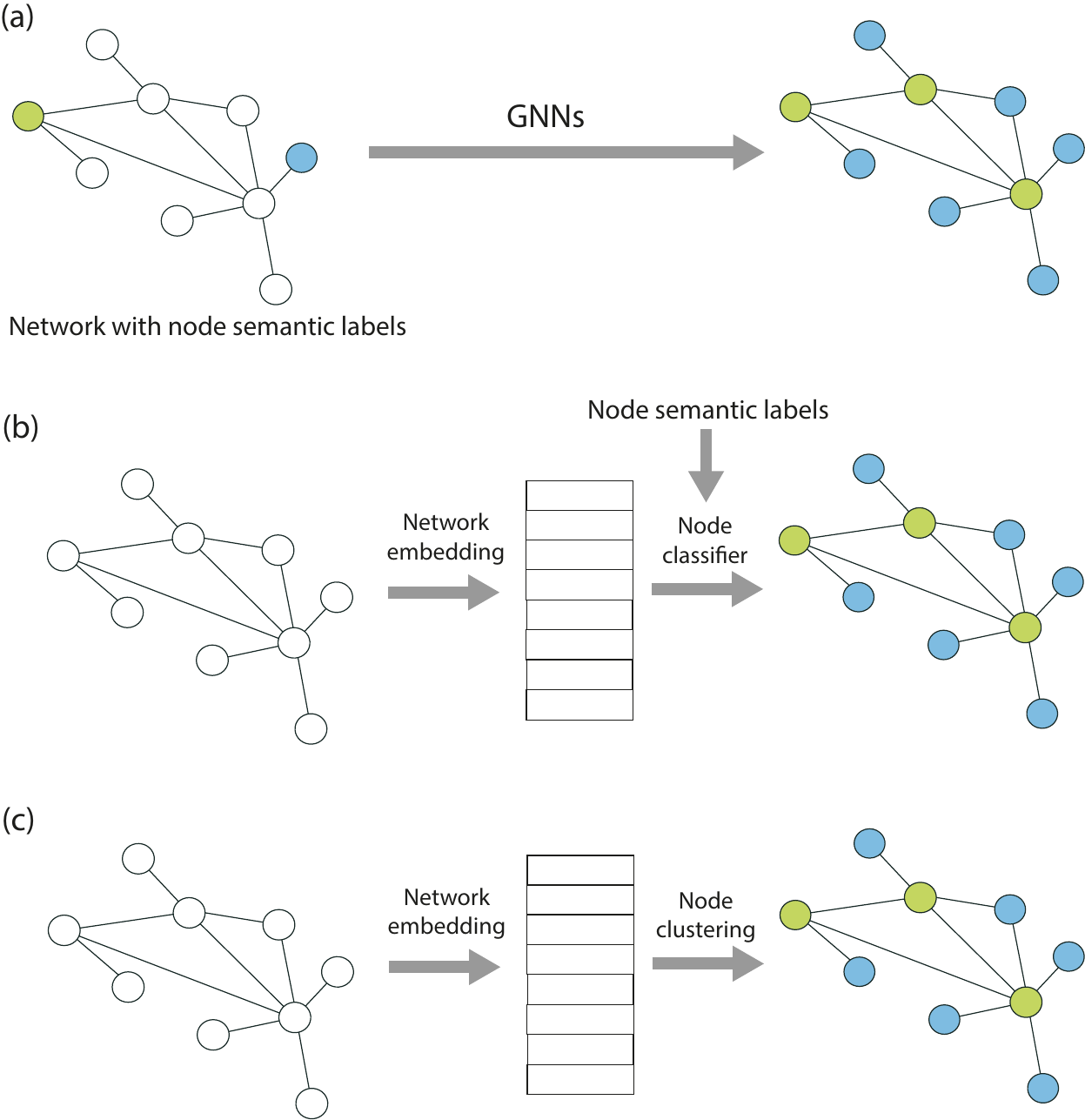}
\caption{Comparison of pipelines for node classification. 
(a) In (semi-)supervised node classification with GNNs, network structure and node semantic labels of nodes in the training set are provided to the model; 
(b) Network embedding obtains low-dimensional representations for nodes without using semantic labels, then a classifier is trained in a semi-supervised fashion using labels of nodes in the training set; 
(c) Most challenging setting in which both network embedding and prediction of node semantic labels do not use supervision from labels in the training set.
}
\label{fig:pipeline_comparison}
% \vspace{-3mm}
\end{figure*}

In Section~\ref{sec:preliminaries}, we gave definition of homophily ratio ($h$), here, in order to provide a better understanding of heterophilous networks, another heterophily ratio is given:
\begin{definition}[Heterophily Ratio $\hat{h}$]
\label{def:heterophily_ratio}
    The heterophily ratio $\hat{h}$ of $\mathcal{G}$  describes the relation between node labels and network structure.
    Similar to the Definition~\ref{def:homophily_ratio}, edge heterophily ($\hat{h}_{\it edge}$) and node heterophily ($\hat{h}_{\it node}$), which can be formulated as
    \begin{equation}
    \label{eq:heterophily}
      \begin{aligned}
      \hat{h}_{\it edge} &= \frac{|\{(u, v): (u, v) \in \mathcal{E} \wedge y_{u} \neq y_{v} \}|}{|\mathcal{E}|} 
      \qquad\qquad
      \hat{h}_{\it node} &= \frac{1}{|\mathcal{V}|}\sum_{v \in \mathcal{V}} \frac{|\{u: u \in \mathcal{N}^{1}_{v} \wedge y_{u} \neq y_{v}\}|}{|\mathcal{N}^{1}_{v}|}
      \end{aligned}
    \end{equation}
    Specifically, where $\hat{h}_{\it edge}$ evaluates the fraction of edges between nodes with different class labels (and $h_{\it edge} + \hat{h}_{\it edge} = 1$);
    $\hat{h}_{\it node}$ evaluates the overall fraction of neighbouring nodes that have different class labels (and $h_{\it node} + \hat{h}_{\it node} = 1$)).
    Figure~\ref{fig:example_hete_network}-(a) shows an example network with $\hat{h}_{\it edge}=\hat{h}_{\it node}=0.8$.
\end{definition}

This paper focuses on the unexplored and challenging research task: unsupervised network embedding with heterophily. 
Therefore, in the experimental evaluation section (Section~\ref{sec:experiments}), we mainly utilise the node clustering task as the downstream application to evaluate the quality of embeddings generated by different models. 
As shown in Figure~\ref{fig:pipeline_comparison}-(c), we do not use any supervisory signal during the pipeline. 
On the contrary, the (semi-)supervised node classification settings (as shown in Figure~\ref{fig:pipeline_comparison}-(a)) takes known node semantic labels to train the model to predict the class label of unknown nodes. 
% 
% Note that we acknowledge that some works adopt a different experiment pipeline that does not use node semantic labels for network embedding but utilises node labels to train a classifier, after obtaining unsupervised node representations, to predict labels of test nodes~\citep{VFHLBH19,PHLZRXH20,BKC21}.
% We argue that this is not a precise setting because if some node semantic labels are available to train a classifier, we can directly utilise them to train the NE model (the typical supervised setting). 
% 
% Therefore, we mainly focus on the node clustering task strictly following the unsupervised setting.
Therefore, we provide results of a number of real-world datasets (Table~\ref{tab:classification_results_real}) that follow the settings~\citet{VFHLBH19,PHLZRXH20,BKC21} to show the generality of SELENE. 
We employ two node classification evaluation metrics: Micro- and Macro-F1 Scores. 
For the node classification task implementation, we adopt the code from official implementation of \citet{BKC21}.
We train a l2 regularised logistic regression classifier from the Scikit learn library\footnote{\url{https://scikit-learn.org/stable/index.html}}. 
We also perform a grid search over the regularisation strength using following values: $\{ 2^{-10}, 2^{-9} \dots, 2^{9}, 2^{10}\}$.

Moreover, we also report the experimental results on another major graph application task, i.e., link prediction, which wants to predict missing links to complete incomplete graph structure. 
The link prediction experimental settings are similar to node2vec paper~\citep{GL16}. 
Particularly, we randomly remove $20\%$ edges from the graph as positive edges, and set half of these edges as validation positive edges and test positive edges, respectively.
We also sample another set of validation and test negative edges, which has the equal number of negative edges (i.e., edges that do not exist in the complete known graph structure) as positive edge set. 
The rest positive edges are used as training set to train the network embedding model to learn node embeddings, the training mechanism is the same as unsupervised network embedding.
In the end, we can predict missing edges by investigating the similarity between a pair of node embeddings. 
We employ ROC-AUC metric to evaluate the link prediction performance. 
For each dataset, we randomly generate $10$ groups of samplings to test each model and report the average values in Table~\ref{tab:link_prediction_results_real}. 
Results in Table~\ref{tab:link_prediction_results_real} show that \textit{(i)} homophily ratio has significant influence on link prediction task.
The performance of network embeddings models on homophilous dataset, i.e., DBLP, is more outstanding than on heterophilous datasets. 
And \textit{(ii)} SELENE performs priority on this task as well. 

\section{Additional Information on Barlow-Twins Loss Function}
\label{sec:appendix_more_about_bt_loss}

Barlow-Twins~\citep{ZJMLD21} is a method that learns data representations using a symmetric network architecture and an empirical cross-correlation based loss function. 
Specifically, given an input image $\mathbf{X}$ and a convolutional neural network models $f_{\theta}$. 
We can compute the image embedding $\mathbf{H} = f_{\theta}(\mathbf{X})$. 
Barlow-Twins loss is formally desined as:
\begin{equation}
\label{eq:lossBarlow}
    \mathcal{L}_{BT} \triangleq  \underbrace{\sum_i  (1-\mathcal{C}_{ii})^2}_\text{invariance term}  + ~~\lambda \underbrace{\sum_{i}\sum_{j \neq i} {\mathcal{C}_{ij}}^2}_\text{redundancy reduction term}
\end{equation}
where $\lambda$ is a positive constant trading off the importance of the first and second terms of the loss, and where $\mathcal{C}$ is the cross-correlation matrix computed between the outputs of the two identical networks along the batch dimension:
\begin{equation}
\label{eq:crosscorr}
    \mathcal{C}_{ij} \triangleq \frac{
    \sum_b \mathbf{H}^A_{b,i} \mathbf{H}^B_{b,j}}
    {\sqrt{\sum_b {(\mathbf{H}^A_{b,i})}^2} \sqrt{\sum_b {(\mathbf{H}^B_{b,j})}^2}}
\end{equation}
where $b$ indexes batch samples and $i,j$ index the vector dimension of the networks' outputs. 
$\mathcal{C}$ is a square matrix with a size  of the dimensionality of the network's output and with values comprised between $-1$ (i.e. perfect anti-correlation) and $1$ (i.e. perfect correlation). 

Intuitively, the \emph{invariance term} of the objective, by trying to equate the diagonal elements of the cross-correlation matrix to 1, makes the embedding invariant to the distortions applied.  The \emph{redundancy reduction term}, by trying to equate the off-diagonal elements of the cross-correlation matrix to 0, decorrelates the different vector components of the embedding. This decorrelation reduces the redundancy between output units so that the output units contain non-redundant information about the sample.

%------------------------------------------------------------------------------

\end{document}